\documentstyle[aps,multicol,epsf]{revtex}
\begin{document}
\draft
\widetext
\title{Electron spin teleportation current through a quantum dot array
operating in the stationary regime}
\author{Olivier Sauret$^a$, Denis Feinberg$^{a}$, Thierry
Martin$^{b}$}
\address{$^a$Laboratoire d'Etudes des Propri\'etes
Electroniques des Solides, Centre National de la Recherche 
Scientifique, BP166, 38042 Grenoble, France}
\address{$^b$
Centre de Physique Th\'eorique et  Universit\'e de la
M\'editerran\'ee, Case 907, 13288 Marseille, France}
\maketitle
\widetext
\begin{abstract}
An electron spin state teleportation scheme is described in detail.
It is based on the protocol by Bennett et al.  [Phys.\ Rev.\ Lett.\ {\bf
70}, 1895 (1993)], and involves the production and detection by
superconductors of entangled pairs of electrons. Quantum
dots filter individual electron transitions, and the whole
teleportation sequence is selected in a
five-dot cell by electrostatic
gating in the stationary regime (no time dependent gate voltages):
i) a normal dot carry the electron spin state to be
teleported, two others carry the ancillary entangled pair; ii) two
superconducting dots,
coupled by a superconducting circuit, control the injection of the source
electron and the detection of the teleported electron.  This teleportation
cell is coupled to emitter and receiver reservoirs.  In a steady
state, a spin-conserving current flows between the reservoirs, most
exclusively carried by the teleportation channel.  This current is
perfectly
correlated to a Cooper pair current flowing in the superconducting circuit,
and which triggers detection of the teleported electron.  This latter
current indeed carries the classical information, which is necessary to
achieve
teleportation.  The average teleportation current is
calculated using the Bloch equations, for weakly coupled spin reservoirs.
A diagnosis of teleportation is proposed using noise correlations.
\end{abstract}
%\begin{multicols}{2}
%\narrowtext

% superconductivity, proximity effect
% single electron tunneling, Coulomb blockade
% entanglement

\section{Introduction}

Teleportation belongs to fundamental science since Bennett and
coworkers proposed a protocol for quantum teleportation of a two-state
particle \cite{bennett}. This means reconstructing the quantum
state of a particle at a distant place, on a preexisting particle which
state was previously undetermined. Of course, any measurement of the
quantum state
to be teleported must be avoided whatsoever. Bennett et al.  proposed to
take advantage of the non-locality of quantum mechanics
\cite{aspect,mandel_zeilinger},
celebrated in the EPR ``paradox'' \cite{EPR}.  To this purpose, a pair of
entangled particles is produced. This means that the state of each of them
is
undetermined, while it is fully determined once a measurement is made on
the other.  One member of the pair is given to the sender, Alice, the other
one (the target) to the receiver, Bob.  Alice also receives a ``source''
particle in an
unknown state, which she wants to teleport to Bob.  Then she performs a
joint measurement on this particle and her
member of the entangled pair, so that as to measure them in an entangled
state.  As a result, the state of the target member of the pair, in the
receiver (Bob)'s
hands, is simultaneously determined.  Alice must send the result of her
measurement as a classical signal.  The state of the source particle can
then be retrieved by Bob, by applying to the target particle a unitary
transformation.  The state of
the source particle has been destroyed during the process (no-cloning
theorem
\cite{zurek}).  An essential point is that, despite the simultaneity of
Alice's measurement
and Bob's particle state projection, teleportation is completed only when
the classical information about the result of the joint measurement (four
possible Bell states thus two classical bits) has been received.
Therefore, as a mean of
transmitting information, TP does not violate any fundamental law.  The
quantum information
stored in the original state (qubit) has been split into a ``quantum''
channel (the entangled pair) and a classical channel.

The first experimental verification of teleportation has been performed
with
photons \cite{bouw}, and was followed by several others
\cite{autretelep}.
Entangled photons with correlated orthogonal
polarizations (antisymmetrical state) were produced by parametric-down
conversion, and measurement of one of the four possible entangled states
(the singlet) was achieved by polarized beam splitters.  This
simplification changes a little, though not fundamentally, the
scenario of Bennett {\it et al.}: the classical signal carries simply a
``yes'' or ``no'' answer
concerning detection, and if the answer is yes, Bob has readily in hands
the original state, without needing any further transformation.  As shown
by
the authors of Ref. \cite{bouw}, this simplification does not affect 
the quantum correlations
of the input and output particles (defined as the fidelity of TP), but it
reduces the
efficiency of TP (success has a probability $1/4$).
The TP protocol is thus rendered slower by a factor $1/4$. In
practice, unambiguous detection of teleportation requires coincidence
measurements of photons at four detectors (one for the detected particle,
two for the Bell measurement and one for a test for emission of the source
particle), and it relies on the optimal control of individual photons
achieved
in modern quantum optics devices.  It is in fact important to keep track of
the emitted pair and the source particle, in order to control that Bob
indeed measures the twin of the photon which experiences the joint
measurement by Alice, and not a member of a previously or a subsequent
emitted pair (in which case no correlation would be expected).

The proposal of Bennett and coworkers and the subsequent experiments in
quantum
optics or atomic physics \cite{bouw,autretelep} and NMR \cite{NMR} 
provide a beautiful
illustration of
the power of entanglement as a basic resource for quantum information
\cite{q_information}.  TP appears as a promising way to send unaltered
quantum information, for instance by entangling fragile qubits with more
robust ones, or with qubits that can be propagated over long distances
(photons for instance).  The same principle allows to swap entanglement
between
successive particles, therefore entangling distant ones.  Other
applications include distributing information among networks, or
error-correcting codes \cite{Bennettcorr}.

Search for scalability naturally leads to a quest for
similar processes in solid-state environments.
The idea of electron/spin transport for quantum information processing
schemes has also been developed in Refs. \cite{larionov_kane}.
Here we shall focus on electron transfer between dots.
In fact the strong
advantages of photons
(weak interactions with the environment, allowing long-distance coherent
propagation)
turn out to be also inconveniences : single or pair photon sources are
weak, and
it is difficult to operate gates on photon ensembles since they interact
weakly,
only through non-linear media. On the other hand, electrons can be
produced one by one,
using Coulomb blockade in quantum dots, or in pairs
\cite{choi_bruder_loss,loss,lesovik_martin_blatter,DF}.

Since electrons are charged particles,
it is in principle possible to operate with a variety of gates on them,
taking advantage of
their Coulomb interactions in nanostructures. Electronic systems also have
obvious
advantages toward integration.
The main drawback of electronic proposals is
that the underlying interactions can also lead to strong decoherence
effects.
Yet, because high intensity single electron sources can be operated, the
relevant time scales
can be very short, and there is hope that  quantum coherence of
individual qubits can be controlled over
distances ranging between microns and millimeters.

Exploring entanglement in the solid state concerns
the practical manipulation of qubits, but also the investigation of
fundamental
phenomena such as a proof of non-locality with massive or fermionic
particles
like electrons \cite{thierry_Bell}.  An existing proposal for TP considers
excitons in coupled quantum boxes, that can be manipulated optically
\cite{excitons}.
Alternatively, an important issue arises when considering the electron spin
degree of freedom, which is a candidate as a qubit for integrated
quantum information devices \cite{loss_divincenzo}.
To produce electron pairs in an entangled spin
state, one needs as a source a device where electron spins are correlated
by
their previous interactions or by their statistics, but the two particles
can be dissociated
while keeping this correlation.  Up to now, proposals have been made using
i) Cooper pairs in superconductors
\cite{DF,choi_bruder_loss,loss,lesovik_martin_blatter} or ii) singlet
states on a discrete level in quantum dots \cite{oliver,saraga}. Entangled
pairs can then be produced by the use of energy
\cite{choi_bruder_loss,loss,lesovik_martin_blatter,oliver,saraga} or spin
filtering \cite{DF,lesovik_martin_blatter}.
Starting from this
elementary unit for entanglement, the construction of more complex devices
relies on the analogy
between photon propagation in waveguides and phase-coherent electron
transport in nanostructures, which have been well illustrated by the
fermion version of the Hanbury-Brown and Twiss intensity correlations
\cite{martin_landauer_buttiker,HBT}.
However, further possibilities are opened by using Coulomb interactions :
transport of electrons, one electron at a time,
can indeed be forced with the help of electrostatic barriers.
Electrons are trapped in quantum dots, separated by tunnel barriers and
their
transitions between dots and reservoirs are controlled by
electrostatic
gates \cite{devoret}. Moreover, as shown in this work, correlations between
transitions of different electrons at different barriers can indeed be used
as a powerful tool to control and manipulate the spin qubits.  Then,
although
time-resolved control of individual electrons is still out of reach,
teleportation can
be achieved in a steady state operation through electrostatic couplings
only \cite{nous}.
For a correct choice of the system parameters, we show in the following
that an elaborate sequence of transitions
can be selected. For instance, in the teleportation process, a strict
control is required on the time sequence: the joint measurement of the
source
particle (1) and of the particle (2) must occur only after the pair (2,3)
is emitted,
and the detection of the ``teleported'' particle (3) must wait for the
joint measurement of
particles (1,2).  As explained in detail in the present paper,
electrostatic couplings between
dots, together with individual gating, provide the necessary correlations
to
filter a unique transition sequence through the cell, without any temporal
gate
control whatsoever.

In the following, a detailed description of the operation of the
teleportation cell is presented.  An overall qualitative presentation is
the
subject of Section II. A full derivation of the microscopic and effective
Hamiltonians is given in Section III, and justifications are provided in
Appendix A.
Section IV contains the discussion of
the electrostatics of the cell and the selection of the teleportation
sequence.  Section V is devoted to the calculation of the steady state
teleportation
current using a Bloch equation approach.  Section VI discusses the overall
similarities
and differences with the quantum optics implementation of teleportation. A
summary of our results, also discussing possible extensions of our solid state
teleportation
proposal, is provided in the conclusion (Section VII).

\section{Qualitative description of the teleportation protocol}

\subsection{Description of the teleportation cell}

\begin{figure} \centerline{\epsfxsize=10cm %\epsfysize=6.5cm
\epsfbox{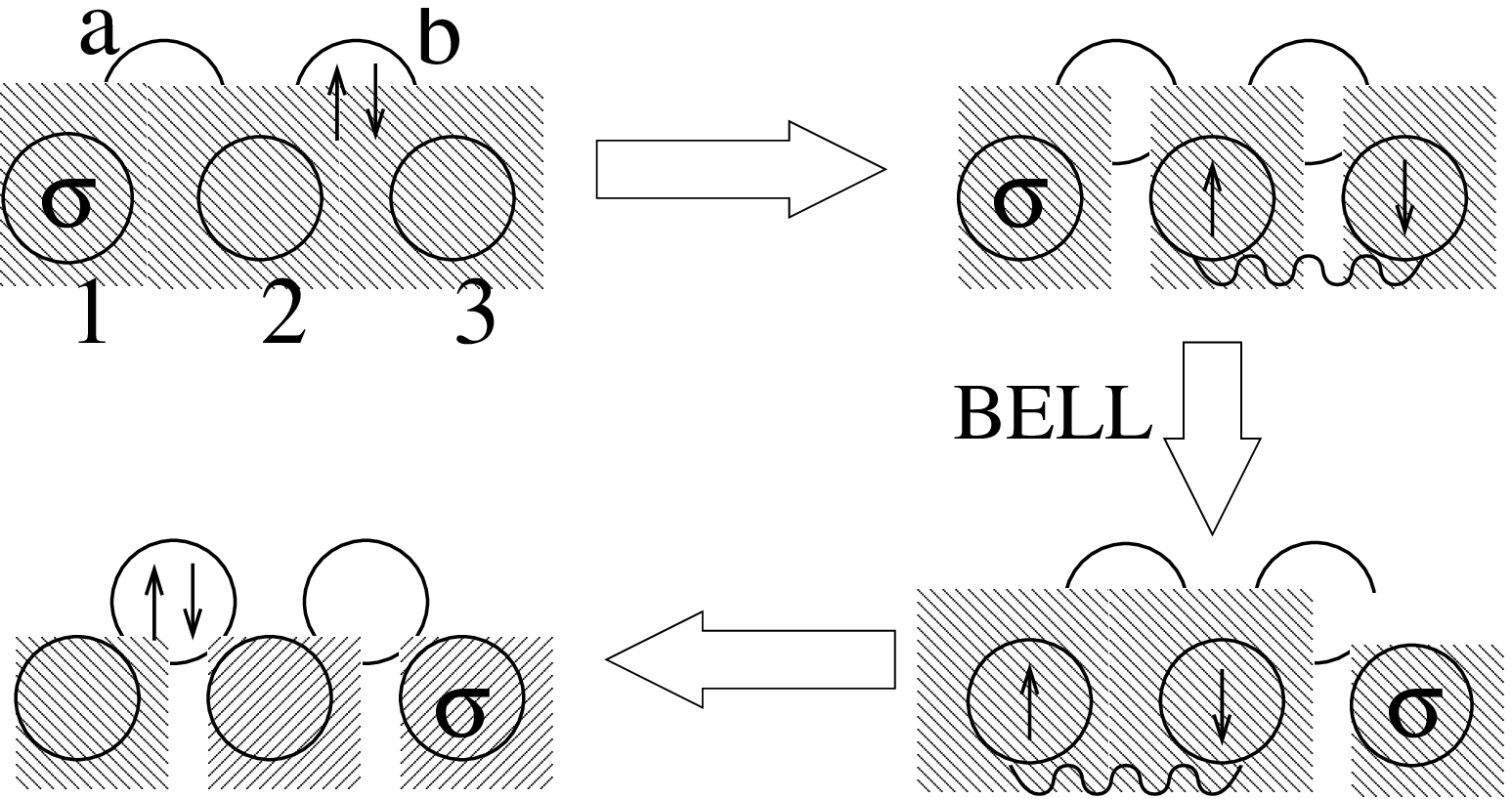}}
%\narrowtext
\vspace{4mm}
\caption{The elementary teleportation sequence. $3$ normal dots (shaded)
labelled $\bf 1,2,3$ can only
accommodate $0$ or $1$  electron, while the $2$ superconducting dots $\bf
a,b$ can accommodate $0$ or
two electrons in a  Cooper pair state. The sequence starts with an
injection process from reservoir $L$,
followed by the Bell decomposition of the $3$ electron wave function in
the normal dots,
and terminates by the detection in $R$.
\label{fig1}} \end{figure}

Let us describe the set-up which allows teleportation of the electron
spin.
The TP cell is defined as comprising the dots $\bf 1,\bf 2,\bf 3,\bf a,\bf
b$ (see Fig. \ref{fig1})
and the circuit $S$ (Fig. \ref{fig2}a).
The basic resource, the entangled pair of electrons, is produced by
superconducting dot $\bf b$ according to the recent proposal
\cite{choi_bruder_loss,loss}: two normal quantum dots (numbered $\bf 
2$ and $\bf
3$)
form neighboring tunnel junctions with a singlet superconducting
electrode.
Electrostatic gates tune the dot chemical potentials such that the
transition
of a Cooper pair from the superconductor to the couple of dots is resonant
if and only if each partner of the pair is sent to a different dot :
conversely,
adding to or subtracting two electrons from a single dot is strongly
suppressed by Coulomb energy.  Due to the symmetry of the Cooper pair
wave function in the superconductor, the two electrons which are added in
dots $\bf
2,\bf 3$ are in the antisymmetric singlet state, which is entangled.  Among
the four Bell states available with two spin $\frac{1}{2}$ particle, none
of the three triplet combinations can be created.  The double
tunneling process involved in the creation of the entangled state is often
denoted as a non-local (or crossed) Andreev process \cite{DF,falci}.  Its
amplitude depends
on the tunneling amplitudes at both junctions (assumed equal for
simplicity),
and on the distance $l$ between junctions
\cite{choi_bruder_loss,falci,epj_feinberg}. The
latter dependence involves exponential decay of quasiparticles in the 
bulk (clean)
superconductor (on the coherence length $\xi_0$) and an
algebraic factor
$(\lambda_F / l)^2$.  Therefore a basic limitation of the ``Andreev
entangler'' is that the distance between the two junctions must not
typically exceed a few nanometers.  Improvement of the algebraic factor
  is obtained by reducing the effective dimensionality
of the superconductor \cite{recher,bouchiat} or using a "dirty"
superconductor \cite{diffusive}.

The source electron is created as an additional charge in dot 1.  The
analyzer, which allows ``Alice'' to perform the joint measurement, is
chosen to be another superconductor.
This possibility of using the same physical
phenomenon both for production and detection of entangled pairs, is
specific to solid state nano-electronics.  This is in
contrast with optics where parametric down conversion produce the pairs
and polarized beam splitters detect antisymmetric pairs in an irreversible
way.  In the present case, having an irreversible transition of
two electrons tunneling from dots $\bf 1$ and $\bf 2$ to superconductor
$\bf a$ (Alice) provides a measurement of those electrons in a singlet
state.  Note that if the electrons were to oscillate between dots 
$\bf 1$ and $\bf 2$
(as a singlet) and dot $\bf a$ (as a Cooper pair) there would be no measurement
whatsoever. Below, we shall see that the irreversibility is
brough the presence of a reservoir which injects an electron in $\bf 1$.
This is analogous to the measurement performed by Alice in
the Innsbruck optics experiment \cite{bouw}.
In this experiment, the two photons which form the singlet state
are destroyed by the measurement process.

If needed, the production of a spin-polarized electron can be achieved by
a spin-polarized
reservoir $\bf L$, connected to dot $\bf 1$ by a tunnel junction.  This
polarized source could be made of a strongly polarized ferromagnet
(half-metal) or based on any other injection scheme, using semiconductors
for instance.  Symmetrically, detection of the teleported electron, sitting
in dot $\bf 3$, can be achieved by a spin-polarized reservoir $\bf R$, in
the
hands of ``Bob'' (Figure 1).

It is convenient to connect the superconducting electrodes $\bf a$ and $\bf
b$ by a superconducting circuit $\bf S$. Indeed, once Alice in $\bf a$
detects a singlet pair, this pair can flow through the circuit $\bf S$
towards $\bf b$, where, as we shall see, it triggers detection by Bob.
In addition, optimum operation of the
device requires to correlate the charge transitions in the normal dots and
superconducting electrodes.  To this purpose, $\bf a$ and $\bf b$ are
chosen to be superconducting dots with sizeable Coulomb charging energy.
This enables to absorb/eject electron pairs one by one.

%%%%%%%%%%%%%%%%%%%%%%%%%%%%%%%%%%%%%%%%%%%%%%%%%%%%%%%%%%
%%%%%%%%%%%%%%%%%%%%%%%%%%%%%%%%%%%%%%%%%%%%%%%%%%%%%%%%%%

\subsection{Bell state decomposition}

Before discussing the conditions on the parameters of this set-up, let us
recall why the measurement on $(\bf 1, \bf 2)$ achieves teleportation
\cite{bennett,bouw}.  Having an electron in dot $\bf 1$ in the state:
\begin{equation}
|\sigma\rangle_1\ = \alpha|\uparrow\rangle_1 + \beta
|\downarrow\rangle_1~,
\end{equation}
an entangled pair of electrons is created in dots
$\bf 2, 3$ in the singlet state $|\Psi^S\rangle_{23}$.
The latter is one of the four basis orthonormal Bell states (one singlet
and
three triplets)
\begin{eqnarray}
|\Psi^S\rangle &=& 2^{-1/2}(|\uparrow\downarrow\rangle -
|\downarrow\uparrow\rangle); ~~~~ |\Psi^{T0}\rangle =
2^{-1/2}(|\uparrow\downarrow\rangle + |\downarrow\uparrow\rangle)
\nonumber\\
|\Psi^{T-}\rangle &=& 2^{-1/2}(|\uparrow\uparrow\rangle - |\downarrow
\downarrow\rangle); ~~~~ |\Psi^{T+}\rangle =
2^{-1/2}(|\uparrow\uparrow\rangle + |\downarrow\downarrow\rangle)
\end{eqnarray}
One verifies that the resulting three-electron state can be
rewritten as
\begin{eqnarray}
|\Psi\rangle_{123} = |\sigma\rangle_1 |\Psi^S\rangle_{23} &=&
-\frac{1}{2}|\Psi^S\rangle_{12} (\alpha|\uparrow\rangle_3 + \beta
|\downarrow\rangle_3) + \frac{1}{2}|\Psi^{T0}\rangle_{12}
(-\alpha|\uparrow\rangle_3 + \beta
|\downarrow\rangle_3)
\nonumber\\
&~&+ \frac{1}{2}|\Psi^{T-}\rangle_{12} (\beta|\uparrow\rangle_3 + \alpha
|\downarrow\rangle_3) + \frac{1}{2}|\Psi^{T+}\rangle_{12}
(-\beta|\uparrow\rangle_3 + \alpha |\downarrow\rangle_3)
\label{bell_state_decomposition}\end{eqnarray}
In the original scenario \cite{bennett}, Alice is able to perform a
measurement
of any Bell state on particles $\bf 1,\bf 2$.  This at the same time
projects the state of particle $\bf 3$ onto a particular state, which Bob
can transform into the original state $|\Psi\rangle_3= |\sigma\rangle_3\ =
\alpha|\uparrow\rangle_3 + \beta |\downarrow\rangle_3$ by applying an
appropriate
unitary transformation.  The latter is known when Alice sends by a
classical
two-bit channel the result of her measurement, which therefore completes
teleportation.  In particular, when the result is the singlet
$|\Psi^S\rangle$, the state in hands of Bob is nothing but the original
state of particle $\bf 1$, up to a minus sign.  Bouwmester et al.
\cite{bouw,bouw2} simplified this protocol : measuring only the state
$|\Psi^S\rangle$ still allows teleportation, provided Alice sends as an
information that she indeed completed the measurement.

The set-up we propose for teleportation of electron spins is similar in
its
principle.  Once three additional electrons are in dots $\bf 1, \bf 2, \bf
3$, the three possible triplet states for electrons in $\bf 1, \bf 2$ lead
to no
detection, while the singlet state can trigger (with probability
$\frac{1}{4}$) an Andreev transition to
the superconducting dot $\bf a$, which acts as Alice's measurement
apparatus.
The remaining spin in dot $\bf 3$ (target) acquires the same state
$\sigma$ as the initial spin in dot $1$, but teleportation is completed
only when the electron is detected in reservoir $\bf R$ (Bob).

\subsection{The teleportation sequence}

The teleportation sequence (see Fig. \ref{fig1}) is the following : having an
additional electron in dot $\bf 1$, an entangled pair is created in dots
$(\bf
2,\bf 3)$.  Thus one electron occupies each of the normal dots. If 
the two electrons
in $(\bf 1,\bf 2)$ are in the singlet state, they can tunnel in dot 
$\bf a$. This
becomes a true measurement when another electron enters dot $\bf 1$ 
from lead $\bf L$,
blocking the singlet pair in dot $\bf a$. At the same time the spin 
state of the electron
  previously in $\bf 1$ is transfered to $\bf 3$, as shown by
Eq. \ref{bell_state_decomposition}.
Let us show how this teleportation sequence is embedded in a teleportation
cycle.
Indeed (see Fig. \ref{fig4}), the pair absorbed in $\bf a$ can flow
from $\bf a$ to $\bf b$ and trigger the transfer and detection of the
teleported
electron in $\bf R$, thus allowing to recover the initial state in Fig.
\ref{fig1}.
It can then serve as source for another entangled pair, for the next
teleportation cycle (Fig. \ref{fig4}).  It is essential to notice that
irreversibility of the pair production
(from $\bf b$) and of the measurement (in $\bf a$) is provided by the
coupling to
the reservoirs : Coulomb repulsion between dots $\bf b$ and $\bf 3$ makes
the `` target '' electron in $\bf 3$
leave towards reservoir $\bf R$,
which in turn lowers the energy of the pair in dot $\bf b$, allowing
resonance with $(\bf 2, \bf 3)$.  Similarly, the pair in $(\bf 1, \bf 2$)
oscillates back and forth to $\bf a$, but it becomes localized in $\bf a$
when a new source electron enters dot $\bf 1$. This in turn
raises the energy in dot $\bf a$ and allows resonance between $\bf a$
and $\bf b$. The above set-up, together with {\it static} gating of dots
$\bf 1,
\bf 2,\bf 3$, allows efficient filtering of single-electron transitions
(from $\bf L$
to $\bf 1$ and $\bf 3$ to $\bf R$) and two-electron transitions (from $\bf
b$ to $(\bf 2,\bf 3)$ and from $(\bf 1,\bf 2)$ to $\bf a$).  But it should
also
prevent spurious transitions : it is indeed essential, to ensure fidelity
of the teleportation process, that the target electron in $\bf 3$ does not
escape
(is detected by Bob) before Alice performs the measurement in $\bf a$.  As
shown in detail in Section IV, this is obtained by appropriately choosing
the gate potentials of each dots.  The Coulomb correlations induced by the
electrostatics of this five-dot system can indeed exactly select the
correct teleportation sequence.

Let us emphasize that transfer of a
charge $2e$ from dot $\bf a$ to dot $\bf b$ plays the role of the
classical channel. Even more, this classical signal strictly conditions
the exit of the target electron.
Therefore, according to the laws of quantum mechanics, teleportation is
achieved without ambiguity, with a fidelity conditioned
by the few spurious processes like cotunneling. This is in contrast
with the optics experiment where (externally operated) time correlations
are used to select the TP events from
spurious ones.

The teleportation process manifests itself as a spin-conserving current
passing from reservoir $\bf L$ to reservoir $\bf R$.
Below we shall consider a situation where the cell
(excluding the leads) is chosen to be symmetric,
an assumption which allows to reduce the number of parameters.
Then, the direction of this
current is fixed by applying a voltage bias $V$ between
$\bf L / \bf 1$ and $\bf 3 / \bf R$.  For each
electron disappearing in $\bf L$ and each electron appearing in $\bf R$
with the same spin, exactly one Cooper pair flows through the circuit $\bf
S$. Yet, in the present electronic device, a full proof of TP -
irrespective of a quantum or
classical description - requires an additional diagnosis, which we discuss
at the end of the paper.

\section{Model}

\subsection{Microscopic Hamiltonian}

The system under consideration (TP cell and reservoirs) is described by
the Hamiltonian:
\begin{equation}
H = H_0 + H_C + H_T
\end{equation}
where $H_0$ contains the energy levels of the isolated elements
(dots and reservoirs),
$H_C$ is the total Coulomb charging energy and $H_T$ is a tunneling
Hamiltonian.  The part $H_0$ reads

\begin{equation}
H_0 = H_1 + H_2 + H_3 + H_a + H_b + H_S + H_L + H_R
\end{equation}
with one-electron energy levels $\varepsilon_{il}$ in the normal dots (in
the
Hartree approximation) ($i = 1,2,3$)
\begin{equation}
H_i = \sum_{l \sigma} \; \varepsilon_{il} \; c^{\dagger}_{il \sigma} \,
c_{il \sigma}
\end{equation}
Normal dots $\bf 1, 2, 3$ must necessarily have a discrete spectrum, in
order to
avoid spin exchange processes which destroy entanglement.
The ``empty'' state corresponds to a state with $2N$ electrons in
the dots, while the ``singly occupied'' state has an odd occupation
number, which means the
addition of one outer electron to the $2N$ electrons
within the dot \cite{loss}. The last doubly occupied level of
dots $\bf 1$ and  $\bf 3$ must lie well below the electro-chemical
potential of reservoirs $\bf L$ and  $\bf R$, so as to prohibit any
spurious transition of ``wrong'' spin electrons between such levels and the
reservoirs.  This sets a lower bound on the level spacing \cite{loss}.

Quasiparticle levels $E_{\alpha k}$ in the superconducting
dots and reservoir $S$ ($\alpha = a,b,S$) enter the Hamiltonian
for the superconductors:
\begin{equation}
H_{\alpha} = \sum_{k \sigma} \; E_{\alpha k} \, \gamma^{\dagger}_{\alpha
k\sigma} \, \gamma_{\alpha k\sigma}
\end{equation}

\noindent
while in normal  reservoirs $N = L,R$

\begin{equation}
H_{N} = \sum_{k \sigma} \; \varepsilon_{N k \sigma} \, c^{\dagger}_{N k
\sigma} \, c_{N k \sigma}
\end{equation}

In the above, the operators $c$ and $c^{\dagger}$ stand for electron
annihilation and creation, and the $\gamma_{k\sigma}$'s are the
usual Bogolubov quasiparticle operators \cite{tinkham} ($\sigma=\pm 1$),
$\gamma_{k \alpha \sigma}=u_{k \alpha}c_{k \alpha \sigma}
+\sigma v_{k \alpha}c^{\dagger}_{-k \alpha -\sigma}$,
with
$u_{k \alpha}=(1/\sqrt{2})(1+\xi_{k \alpha}/E_{k \alpha})^{1/2}$,
$v_{k}=(1/\sqrt{2})(1-\xi_{k \alpha}/E_{k \alpha})^{1/2}$,
$\xi_{k \alpha}=\varepsilon_{k \alpha}-\mu_{\alpha}$ and
$E_{\alpha k}=\sqrt{\xi^{2}_{k \alpha}+\Delta_{\alpha}^{2}}$. Here
$\Delta_{\alpha}$ is the gap in superconductors $\alpha = \bf a, \bf
b$, and $\bf S$ where the chemical potential is set to zero with respect
to the normal reservoirs $\bf L, \bf R$.

One assumes a free-electron spectrum in reservoirs $L,R$ with
spin-dependent quasiparticle energies. To a good approximation,
a continuous BCS-like quasiparticle spectrum in reservoir $S$ and dots
$a,b$
is specified: this requires that the superconducting dots are ``large''
enough.
Finally a discrete (and spin-independent) spectrum is chosen in dots
$1,2,3$.

The charging
Hamiltonian can be written in a compact form as a function of the
occupation numbers
$N_{\mu}=\sum_{\mu l}c^{\dagger}_{\mu l\sigma}c_{\mu l\sigma}$
of the dots ($\mu,\nu=1,2,3,a,b$) \cite{devoret,beenakker}

\begin{equation}
H_C = (1/2) \,\sum_{\mu,\nu} \; C_{\mu\nu}^{-1} \,
(N_{\mu}e-Q_{\mu}) \, (N_{\nu}e-Q_{\nu})
\label{coulomb}
\end{equation}

\noindent
where the $Q_{\mu}$'s are the effective occupation number imposed by the
gate
and bias voltages (see Section IV and Appendix A).  The couplings
$C_{\mu\nu}^{-1}$ are the inverse capacitance matrix elements of the
system.

Here the Coulomb charging energy of the superconducting dots is chosen to
be
smaller than their superconducting gap $e^2/C_{a,b}\ll \Delta_{a,b}$, so
as to
prohibit single electron transitions.

%%%%%%%%%%%%%%%%%%%%%%%%%%%%%%%%%%%%%%%%%%%%%%%%%%%%%%%%%%
\begin{figure} \centerline{\epsfxsize=10cm %\epsfysize=6.5cm
\epsfbox{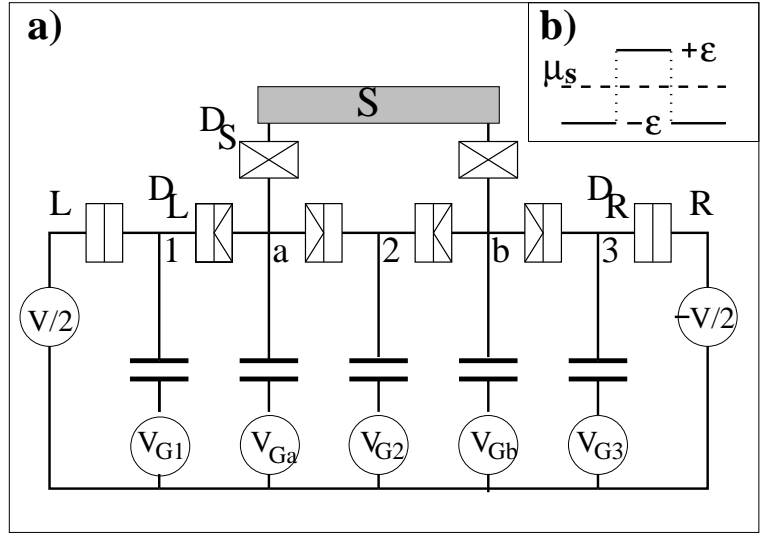}}
%\narrowtext
\vspace{4mm}
\caption{a) The TP cell contains: i)
NN junctions between reservoirs $\bf L,\bf R$ and dots $\bf 1$ and $\bf 3
$; ii)
N-S junctions between $({\bf 1},{\bf a})$, $({\bf a},{\bf 2})$, $({\bf
2},{\bf b})$ and $({\bf b},{\bf 3})$, and S-S junctions between $\bf a$
($\bf b$) and the bulk superconductor $\cal{S}$. Detectors $D_{L,R,S}$
signal the passage of an electron/Cooper pair
in the normal/superconducting circuit. b) Effective energy
level configuration (from left to right) of dots $\bf 1$, $\bf 2$ and $\bf 3$
($\mu_S$ is the superconductor chemical potential ).
\label{fig2}} \end{figure}
%%%%%%%%%%%%%%%%%%%%%%%%%%%%%%%%%%%%%%%%%%%%%%%%%%%%%%%%%%

Let us assume that electronic transitions involve only one level of the
normal dots. This is justified at low enough temperature by the discrete
spectrum of N-dots,
and by the fact that gate voltages can be chosen so as to select only
two possible charge states. Tunneling is supposed to occur only between :
reservoir/normal dot
junctions $\bf L / \bf 1$, $\bf R / \bf 3$, normal dot/superconducting
dot junctions $\bf 1 / \bf a$, $\bf 2 / \bf a$, $\bf 2 / \bf b$, $\bf
3 / \bf b$, and superconducting reservoir/superconducting dot junctions
$\bf S / \bf a$, $\bf S / \bf b$. The one-electron tunneling
Hamiltonian can then be written as
\begin{equation}
H_{T}=H_{L1} + H_{R3} +  H_{Sa} +
H_{Sb} + H_{a1} + H_{a2} + H_{b2} + H_{b3}
\end{equation}
with
\begin{eqnarray}
\nonumber
H_{T} = \sum_{k \sigma} t_{Lk} c^{\dagger}_{Lk\sigma} c_{1 \sigma} +
  \sum_{k \sigma} t_{Rk} c^{\dagger}_{Rk\sigma} c_{3 \sigma} +
  \sum_{kk' \sigma} t_{Sakk'}
c^{\dagger}_{Sk \sigma} c_{ak'\sigma} +
\sum_{kk' \sigma} t_{Sbkk'}
c^{\dagger}_{Sk \sigma} c_{bk'\sigma} \\
+ \sum_{k \sigma} t_{a1k} c^{\dagger}_{ak \sigma} c_{1 \sigma} +
\sum_{k \sigma} t_{a2k} c^{\dagger}_{ak \sigma} c_{2 \sigma} +
  \sum_{k \sigma} t_{b2k} c^{\dagger}_{bk \sigma} c_{2 \sigma} +
  \sum_{k \sigma} t_{b3k} c^{\dagger}_{bk \sigma} c_{3 \sigma} + H. c.
\label{tunneling}
\end{eqnarray}

Notice that no direct tunneling occurs between normal dots, instead,
dot $\bf 2$ is connected to both $\bf a$ and $\bf b$ dots.
Tunneling through the junctions is controlled by the
rates $\Gamma_{L(R)\sigma} = 2\pi t_{L1(R3)}^{2}\nu_{L(R)\sigma}(0)$,
$\Gamma_{S \alpha} = 2\pi t_{S \alpha}^{2}\nu_{S}(0) = \Gamma_{S}$
($\alpha = a,b$) and
$\Gamma_{\alpha i} = 2\pi t_{\alpha i}^{2}\nu_{S}(0) = \Gamma_{\alpha}$
where $\nu_{L(R)\sigma}(0)$ is the spin-dependent density of states at
the Fermi level in reservoirs $\bf L,\bf R$ and $\nu_{S}(0)$
the normal density of states in the superconducting reservoir.

An important constraint concerns the coupling between normal dots and the
reservoirs.  In order to avoid spin exchange between dots
$\bf 1, \bf 3$ and reservoirs, $\Gamma_L$ ($\Gamma_R$)
must be smaller than the superconducting gap, the charging energy and the
chemical potentials
$\mu_L$, $\mu_R$ \cite{loss}.  Moreover, the level spacings in each normal
dot must be larger than $\mu_L$, $\mu_R$ (measured with respect to $\mu_S=0$).

\subsection{Transport processes in the teleportation cell}

The microscopic Hamiltonian of the preceding section
allows for several collective electron transfer processes.
Some of these will be more favorable because of the
intermediate states which they involve (generation of one or more
quasiparticles in the superconducting elements). In particular,
we will be interested in processes where a minimum of
quasiparticles are excited in the superconducting elements
of the teleportation cell.

\begin{figure} \centerline{\epsfxsize=10cm %\epsfysize=6.5cm
\epsfbox{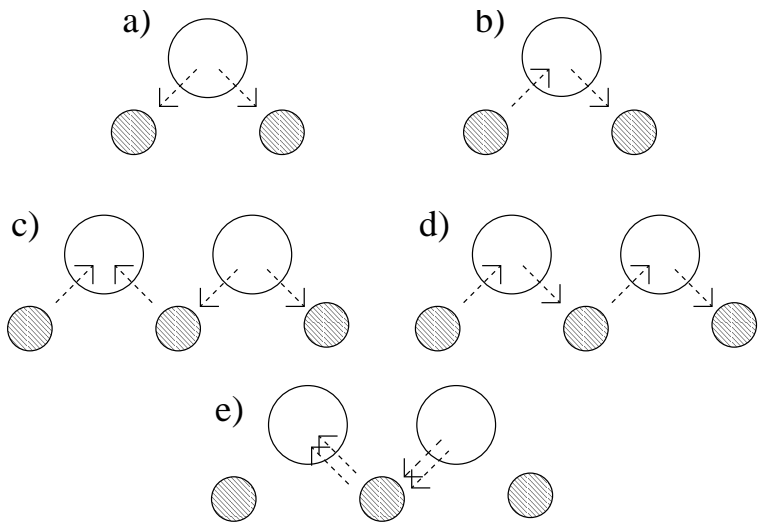}}
%\narrowtext
\vspace{4mm}
\caption{Transport processes between normal quantum dots and
superconducting dots:
a) Crossed Andreev reflection; b) cotunneling; c) teleportation; d)
sequence of two cotunneling
events; e) Josephson tunneling through the central dot.
\label{fig3}} \end{figure}

First consider processes involving two electrons, being transfered
from/to normal quantum dots and a superconducting dot.
These are depicted in Fig. \ref{fig3}.
Crossed Andreev reflection \cite{choi_bruder_loss,falci} --
also called pair tunneling -- (Fig. \ref{fig3}a)
involves the quasi-simultaneous tunneling of two electrons to/from
a superconducting dot to two neighboring dots. As the first
electron tunnels it generates a quasiparticle, which is
then destroyed by the second electron which is transfered.
Cotunneling \cite{cotunneling} describes the transfer of an electron from
one
normal quantum dot to another (Fig. \ref{fig3}b),
via the superconductor (the order of the sequence of the two
tunneling events is arbitrary). The intermediate
state also implies the creation of a quasiparticle, which is
subsequently destroyed as in pair tunneling, and cotunneling is
a spin preserving transition.

%%%%%%%%%%%%%%%%%%%%%%%%%%%%%%%%%%%%%%%%%%%%%%%%%%%%%%%%%%
%%%%%%%%%%%%%%%%%%%%%%%%%%%%%%%%%%%%%%%%%%%%%%%%%%%%%%%%%%

Next, there are three types of processes which involve $4$ single electron
jumps.
Teleportation (Fig. \ref{fig3}c) requires that dot $\bf 1$ is occupied
initially,
and dot $\bf 3$ is occupied in the final state. A Cooper pair escapes from
$\bf b$
and another one is absorbed in $\bf a$. At no step in the process
a directed matter transfer occurs between dot $\bf 1$ and dot $\bf 3$.
This is the opposite case for the succession of two cotunneling
events (Fig. \ref{fig3}d), where, conversely, no Cooper pair transfer
occurs
simultaneously, but an electron is transfered for $\bf 1$ to $\bf 3$
nevertheless.

Note that another possible cotunneling event from $\bf 1$ to $\bf 3$
can involve the bulk superconductor instead, by successive quasiparticle
excitations in S-dot $\bf a$, $\cal S$, and S-dot $\bf b$. However,
this process is negligible: it involves a geometrical factor corresponding to
the propagation of unpaired electrons between $\bf a$ and $\bf b$ through the
bulk superconductor.

One may also consider the
transfer of a Cooper pair from $\bf b$ to $\bf a$, this time
without the transfer of an electron from  $\bf 1$ to $\bf 3$, which can be
viewed as a kind of Josephson coupling between
$\bf b$ and $\bf a$ \cite{choi_bruder_loss,glazman}.
Note that both the successive
cotunneling and the Josephson-like processes
compete with the Andreev processes which are essential for teleportation.

Here, we argue that optimum conditions for the operation of the
teleportation cell
are met when the Andreev pair transitions (from $\bf b$ to ($\bf 2, \bf
3$)
and from ($\bf 1, \bf 2$) to $\bf a$) are chosen to be resonant.
The resonance condition also applies for the Cooper pair
transitions between $\bf a, \bf b$ and the circuit $\bf S$.
This Cooper pair transfer amplitude involves two Josephson
junctions, each junction involving a superconducting dot
(${\bf a}$ or $\bf b$) and the bulk superconductor $\cal S$.
The Josephson coupling associated with each junction
has been computed in Ref. \cite{matveev} starting from
a microscopic, single electron hopping Hamiltonian. Surprisingly,
these Josephson couplings can be reinforced by the Coulomb
blockade effects in the superconducting dots.
In practice, such resonant conditions can be achieved because the
individual levels of dots $\bf 1, 3$ (equal by our choice of a
symmetric cell) and dot $\bf
2$ can be independently chosen, provided they stay within the
superconducting gaps of superconducting dots $\bf a, \bf b$.

In Appendix \ref{tmatrix}, a
perturbative argument is provided
to  show that direct transitions of electrons from $\bf 1$ to $\bf 3$ via
$\bf 2$
can be strongly reduced compared to the teleportation process. The
amplitude of cotunneling being comparable to that of Andreev transitions
\cite{falci}, the way to suppress cotunneling is by raising the level in
$\bf 2$ with respect to $\bf 1, \bf 3$ (Fig. \ref{fig2}b). Then,
contrarily to the
teleportation process, cotunneling involves only non-resonant transitions
and
can be safely neglected to lowest order.
We also show show in Appendix \ref{tmatrix} under which conditions the
Josephson process
can be neglected.

Gate voltages on the five dots of the cell can be tuned such that all the
singlet
pair transitions in the TP cell are either resonant or blocked by Coulomb
interactions (at $T=0$ and neglecting cotunneling). First, pair
transitions between
dot $\bf a$ ($\bf b$) and circuit $\bf S$ have amplitude $T_{J}^{a,b}$.
Second, Andreev pair transitions between dot $\bf a$ and the pair of dots
($\bf 1,\bf
2$), and between dot $\bf b$ and the pair of dots ($\bf 2,\bf
3$), have amplitudes $T_{A}^{a,b}$. The resonant condition for these
amplitudes will be explicit when writing the coherence terms of the Bloch
equations for the reduced density matrix elements.
On the opposite, Cooper pair transitions from $\bf a$ or $\bf b$ to one
of individual dots $\bf 1,\bf 2,\bf 3$ are strongly suppressed by Coulomb
repulsion.
These last
assumptions were also considered in \cite{choi_bruder_loss,loss}, where the
system formed by a
superconductor and two normal dots was shown to behave as a source of
entangled Cooper pairs.  Non-local
(or crossed) Andreev process remains unchanged if
the bulk superconductor is a small superconducting island with
sizeable Coulomb energy, especially in the resonant case.

We further stress that here the driving force for teleportation is the
voltage bias applied to the teleportation cell.  The resulting Cooper pair
current flowing in the superconducting branch is not a Josephson current,
but is instead an Andreev current dragged by the spin-polarized 
current flowing under the effect
of the bias $V$.

\subsection{Effective Hamiltonian}

>From what precedes one can derive an effective tunnel Hamiltonian, which
involves only pair tunneling within the TP cell and single-electron
tunneling to and from the normal reservoirs,

\begin{equation}
H_T^{eff} = H_{L1} + H_{R3} + H_{P}
\label{eff}
\end{equation}

with

\begin{equation}
H_{P} = T_J (\Psi^{\dagger}_a +
\Psi^{\dagger}_b)\Psi_{S}+ T_A^a \Psi^{\dagger}_{12} \Psi_a + T_A^b
\Psi^{\dagger}_{23}\Psi_b + H.c.
\label{pair}
\end{equation}

where the $\Psi_{ij}$ destroys a singlet pair in two
N-dots and $\Psi_{a,b,{S}}$ destroys a Cooper pair in the
superconducting elements. This Hamiltonian will be used in Section V to
derive the average
current through the cell.

\section{Electrostatics of the teleportation cell}

Let us now discuss the central issue of the Coulomb energy balance in
the TP cell. The external variables are the bias voltage $V$ applied
between $\bf L$ and $\bf R$ and the gate voltages $V_{g \mu}$ applied
to dots $\mu = 1,2,3,a,b$. For the sake of simplicity, the TP cell is
assumed to be symmetric, which means : i) the equality of the tunneling
matrix
elements between dot pairs $\bf a \bf 1(\bf2), \bf b \bf 2(\bf 3)$ and ii)
the equality of the capacitances $C_{1a} = C_{2a} = C_{2b} = C_{3b} = C$,
$C_{aS} = C_{bS} = C_{s}$.  Notice that the tunneling amplitudes from
reservoir $\bf L$ or to reservoir $\bf R$ can instead be different: the
symmetry
of the device is only assumed within the teleportation cell.  One
assumes in addition that $C_{L1} = C_{R3} = C_{r}$, and that the gate
capacitances are all equal to $C_{g}$, which is taken much smaller than
$C,C_{s},C_{r}$.
In this section's applications, the teleportation regime will be studied
under the
assumption $C = C_r = C_s$ in order to reduce the number of
parameters.
At temperatures $k_BT << |V|$, the direction of the
current through the cell is determined by the sign of $V$, from left ($\bf
L$) to right ($\bf R$) as a convention.  The dot occupation numbers
$N_{\mu}$ are defined by the excess charge numbers $n_{\mu} = N_{\mu} -
N_{\mu}^0$ with respect to a reference state with even occupancies.
External voltages can be tuned such as the relevant numbers
are $n_{\mu} = 0,1$ for N-dots, and $n_{\mu} = 0,2$ for S-dots, apart from
intermediate quasiparticle states involved in pair transitions, which
involve odd charge states.  This defines the cell configurations ($n_1 \,
n_a \, n_2 \, n_b \, n_3$).  Incorporating the charge numbers $N_{\mu}^0$
into the definition of the effective charges $Q_{\mu}$, one writes the
total
electrostatic energy of the cell

\begin{equation}
E_{n_{1} n_{a} n_{2} n_{b} n_{3}}=
\frac{1}{2}\sum_{\mu,\nu\in{dot}} C_{\mu\nu}^{-1}(n_{\mu}e-Q_{\mu})
(n_{\nu}e-Q_{\nu}))-\frac{(N_L+N_R)eV}{2}
\label{energy_of_config}
\end{equation}

where $Q_{\mu}=C_{g}V_{g{\mu}}$ ($\mu = \bf a, \bf 2, \bf b$) and
$Q_{1}=C_{g}V_{g1} + C_r\frac{V}{2}$, $Q_{3}=C_{g}V_{g3} -
C_r\frac{V}{2}$.
$N_{L}$ and $N_{R}$ are the total number of charges that have passed
through junctions $\bf L/\bf 1$ and $\bf 3 / \bf R$.  The voltage drops at
these junctions are equal to $V/2$ due to the symmetry in the 
junction capacitances.

The electronic transitions within
or out of the cell involve total Coulomb energy differences $\Delta
E_{i}^{f}=\Delta
E_{(n_{1}n_{a}n_{2}n_{b}n_{3})_i}^{(n_{1}n_{a}n_{2}n_{b}n_{3})_f} +
\Delta \varepsilon_{{\mu}}$ between initial and final states.
The second term accounts for the
discreteness of the normal dot spectrum, and will be neglected
compared to the main Coulomb contribution.
Among all possible transitions in the system, several processes can
result in a current from reservoir $\bf L$ to reservoir $\bf R$.  Some of
them involve teleportation of the spin state present in $\bf 1$, others
not.
Selection of the former processes can be achieved at low enough
temperature if their
energy balance is negative, while that of unwanted processes is positive.
Also, spurious processes leading to spin exchange with the reservoirs can
be suppressed in this way.
%%%%%%%%%%%%%%%%%%%%%%%%%%%%%%%%%%%%%%%%%%%%%%%%%%%%%%%%%%%%%%%%%%%%%%%%%%%%%%%%%%%%%%%%%%%%%%

\begin{figure}
\centerline{\epsfxsize=8cm
\epsfbox{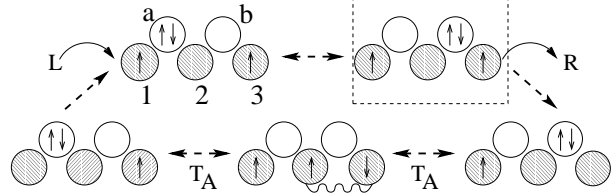}}
%\narrowtext
\vspace{4mm}
\caption{
TP cycle, which operates with a DC bias.
``Horizontal'' transitions only are
resonant. Starting from the framed configuration (upper right), an electron
in $\bf 3$ escapes in $\bf R$; next, a pair (from $\bf b$) creates
an entangled state ${\bf 2},{\bf 3}$ (wiggly line) with rate $T_A$,
leaving all N-dots filled.  A pair ${\bf 1},{\bf 2}$ then escapes in ${\bf
a}$.
The electron in ${\bf 3}$ acquires the spin state of dot ${\bf 1}$,
as confirmed
by the absorption of a singlet state in ${\bf a}$ and the subsequent
injection of an electron from ${\bf L}$.
\label{fig4}}  \end{figure}
%%%%%%%%%%%%%%%%%%%%%%%%%%%%%%%%%%%%%%%%%%%%%%%%%%%%%%%%%%%%%%%%%%%%%%%%%%%%%%%%%%%%%%%%%%%%%%%%

In all the above, it is crucial to have a finite charging energy in 
the superconducting dots.
Transitions involving dot occupation by many Cooper pairs would instead follow,
leading to further spurious processes.
This charging energy is necessary to enforce the desired sequence of events.

\subsection{Allowed transitions within the teleportation cycle}

Allowed transitions are represented in Fig.
\ref{fig4}. The  sequence may (arbitrarily) start
with the evacuation of dot $\bf 3$ triggered by a Cooper pair
in $\bf b$ (see framed section). The electron measured in $\bf R$ 
corresponds to
a previously teleported state.
At the same time, the next electron
to be teleported is already sitting in dot $\bf 1$. This allows next
to deposit a singlet in dots $\bf b$ and $\bf c$ and to
ultimately perform the Bell projection, coming back to the
initial state.
Transitions which do not involve reservoirs are taken
to be resonant, while the injection and detection steps from/to the
reservoirs are taken
to be irreversible.
First, the conditions for resonance
between configurations ($10020$), ($10101$) and ($02001$) can be deduced
from
Eq. (\ref{energy_of_config}), e.g. $\Delta
E_{10020}^{10101}=\Delta E_{10101}^{02001}=0$ :

\begin{equation}\label{andreevb}
Q_1+5Q_2+4Q_3+2Q_a-7Q_b=-\frac{3}{2}e
\end{equation}
\begin{equation}\label{andreeva}
4Q_1+5Q_2+Q_3-7Q_a+2Q_b=-\frac{3}{2}e
\end{equation}

Similarly, the resonance condition between configurations ($12001$),
  ($10001$) and ($10021$), e.  g. $\Delta E_{10001}^{12001}=\Delta
  E_{10021}^{10001}=0$ leads to

\begin{equation}\label{cooperb}
Q_1+5Q_2+4Q_3+2Q_a+8Q_b=13e
\end{equation}
\begin{equation}\label{coopera}
4Q_1+5Q_2+Q_3+8Q_a+2Q_b=13e
\end{equation}

These equations can be simplified, assuming $Q_1 = Q_3$ and $Q_a = Q_b$,
which
yields
\begin{equation}
\label{reson}
Q_1+Q_2=\frac{2}{3}e, \; \; Q_a=\frac{29}{30}e
\end{equation}

Notice that the second condition means for S-dots a quasi-resonance
between states
differing by one Cooper pair.
One then finds, as a simple result, that the above resonance conditions
for
injection or detection processes fix the background
charge on superconducting dots, and relates those on normal dots.  This
still leaves one more gate voltage as a free parameter.

Let us turn to the ``transport'' conditions allowing electronic
transitions from $\bf L$ to
$\bf 1$, and from $\bf3$ to $\bf R$ to occur (assuming $eV > 0$). Two
cases must be distinguished,
depending of the relative energies of
the ``even'' (resonant) states ${(12001),(10001),(10021)}$ compared to the
``odd'' (resonant) states
${(10020),(10101),(02001)}$. First, assume that the even states are more
stable
(which corresponds to $Q_1 > \frac{9}{10}$).
Then the transport conditions can be written as

\begin{equation}
e\frac{V}{2} > \Delta E_{02001}^{12001}, \; \Delta E_{10021}^{10020}
\end{equation}

\noindent
which in the present case result in

\begin{equation}
\label{injdec}
  Q_1 - \frac{9}{10}  < \frac{CV}{e}
\end{equation}

On the other hand, in the other case where ``odd'' states are more stable
than ``even'' ones
($Q_1 < \frac{9}{10}$), one gets similarly

\begin{equation}
\label{injdec2}
-Q_1 + \frac{9}{10} \, < \frac{CV}{e}
\end{equation}

The conditions of Eqs (\ref{reson}) and (\ref{injdec}) are necessary
for the wanted sequence to occur,
but still do not prevent spurious transitions from occurring.
We now consider all possible
processes leaving the sequence states.

\subsection{Unwanted processes}

As discussed earlier, the microscopic Hamiltonian allows for
a number of unwanted processes. These can be minimized or reduced for
physical reasons, keeping in mind that we can tune the energy levels
in the dots.

The first category of unwanted processes concerns the instant at which
charges are injected into $\bf 1$ or detected from $\bf 3$.  According to
the general TP protocol, a singlet state has to be measured in dots ($\bf
1,\bf 2$), by an irreversible transition to dot $\bf a$.  This means that
injection must occur in state ($02001$), to the exclusion of any other
state of the sequence.  This is enforced by forbidding the occupation
number $2$ on dot $\bf 1$, e.g.
\begin{equation}
\label{wronginject}
e\frac{V}{2} < \Delta E_{1\bf{2}0\bf{0}1}^{2\bf{2}0\bf{0}1}, \, \Delta
E_{1\bf{0}0\bf{0}1}^{2\bf{0}0\bf{0}1}, \,
\Delta E_{1\bf{0}0\bf{2}1}^{2\bf{0}0\bf{2}1}, \, \Delta
E_{1\bf{0}0\bf{2}0}^{2\bf{0}0\bf{2}0},
\Delta E_{1\bf{0}1\bf{0}1}^{2\bf{0}1\bf{0}1}
\end{equation}

On the other hand, detection from $\bf 3$ must wait for a classical signal
to be
sent, that the above measurement has been completed.  Since the pair in
$\bf a$, once created, can resonate with $\bf b$ through $\bf S$, the
signal is nothing but the appearance of a pair in $\bf b$, which triggers
detection from $\bf 3$ by means of the Coulomb repulsion.  For this one
must avoid spurious transitions from ($1\bf{2}0\bf{0}1$),
($1\bf{0}0\bf{0}1$) and ($1\bf{0}1\bf{0}1$),
which implies

\begin{equation}
\label{wrongdetect}
e\frac{V}{2} < \Delta E_{12001}^{12000}, \, \Delta E_{10001}^{10000},
  \, \Delta E_{10101}^{10100}, \Delta E_{02001}^{02000}
\end{equation}

The compatibility of Eq.  (\ref{wrongdetect}) with Eq. (\ref{injdec})
requires to satisfy the following conditions

\begin{equation}
\Delta E_{12001}^{12000}, \, \Delta E_{10001}^{10000}, \,
\Delta E_{10101}^{10100}, \, \Delta E_{02001}^{02000} \, > \, \Delta
E_{10021}^{10020}
\end{equation}

These conditions can be shown to be equivalent to the following
set of inequalities between inverse capacitances

\begin{eqnarray}
\nonumber
C^{-1}_{b3}-C^{-1}_{a3} > 0,&&~~~~  C^{-1}_{b3} > 0\\
2C^{-1}_{b3}-C^{-1}_{23} > 0,&&~~~~  2(C^{-1}_{b3}-C^{-1}_{a3}) +
C^{-1}_{13} > 0
\end{eqnarray}

which are always fulfilled, owing to the concavity of the
decrease of the inverse capacitance $C^{-1}_{ij}$ with the distance
$|i-j|$.

An especially important condition concerns the cotunneling process from
dot $\bf 1$ to dot $\bf 2$ (see Section II)
which involves the possible transitions $(12001) \rightarrow
(02101)$,$(10021) \rightarrow (00121)$, $(10001) \rightarrow (00101)$.
These transitions involve positive energies provided that $Q_1 -
\frac{8}{15} > 0$. A sufficient condition to minimize
cotunneling processes is therefore

\begin{equation}
  Q_1 - \frac{8}{15} >> \gamma_a
\end{equation}

Other forbidden transitions concern injection from reservoir $\bf R$, and
detection to reservoir $\bf L$, processes which would contribute to an
``inverse'' current
flowing from $\bf R$ to $\bf L$. One can check that this involves the
condition

\begin{equation}
\frac{11}{10}-Q_1 > \frac{CV}{e}\; \; \; \; \; Q_1 - \frac{1}{6} >
\frac{CV}{e}
\end{equation}

One could also worry about a process transferring an electron from $\bf 1$
to $\bf 3$ accompanied by a net Cooper pair transfer
from $\bf b$ to $\bf a$, but the latter happening in reverse order: the
absorption of the singlet in $\bf a$ would occur before
the creation of the entangled pair in $\bf b$. Its sequence is $(10020)
\rightarrow (02\bar{1}20) \rightarrow (02001)$, e. g.
the absorption of the singlet in $\bf a$ occurs before creation of the
entangled pair in $\bf b$. This process is however
unlikely due to the charge state $-1$ in dot $\bf 2$, which renders the
crossed Andreev transition non-resonant.
%%%%%%%%%%%%%%%%%%%%%%%%%%%%%%%%%%%%%%%%%%%%%%%%%%%%%%%%%%
\begin{figure} \centerline{\epsfxsize=6cm %\epsfysize=6.5cm
\epsfbox{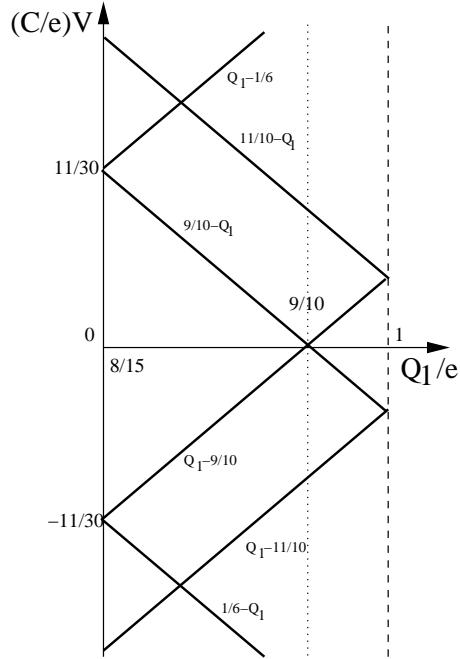}}
%\narrowtext
\vspace{4mm}
\caption{Stability diagram for the teleportation cycle. The two 
tilted rectangles
indicate the parameter domain for the
voltage bias and for the background charge of dot $\bf 1$ which are
required to stay
in the teleportation cycle.
\label{fig5}}
\end{figure}
%%%%%%%%%%%%%%%%%%%%%%%%%%%%%%%%%%%%%%%%%%%%%%%%%%%%%%%%%%
The operating conditions discussed above can be summarized in a  stability
diagram (Fig. \ref{fig5}), plotted as a function of
$Q_1$ and $V$. The two symmetric rectangles tilted at $\pm \pi/4$ contain
the zero
temperature working regimes. Outside the vertical lines on the left and on
the right, cotunneling is favored.
This diagram shows that all the conditions
meet in a relatively large portion of the parameter range. Remarkably
enough, apart from the ``even-odd''
  degeneracy point at $Q_1/e=9/10$,
there is a threshold voltage for teleportation. This means than the system
does no work in the linear regime but instead there
is a ``Coulomb gap'' equal to the energy difference between the even and
the odd states participating to the TP sequence. Notice
that $V$ is also bounded from above, to avoid spurious transitions during
the sequence. The size of the working region guarantees that the 
system can work in presence of
weak thermal and quantum fluctuations. Indeed, the relevant energy 
scale is a sizeable fraction
of the Coulomb charging energy of the dots.

Note that the simplification which consists of choosing the junction
capacitances to be equal is by no way restrictive. It can be shown 
that all the above
results hold in general, the only condition being on the relative values
of the capacitances $C_s$ and $C_r$.

\section{Calculation of the teleportation current}

This Section presents the dynamics of the five-dot teleportation cell,
described
in the above working regime, at zero temperature.
The method is that of the master equation describing the dynamics of
the reduced density matrix $(\sigma_{\mu \nu})$ where the
$\mu$'s are the configuration states retained in the sequence.
A microscopic derivation of the master equation for a system of
dots, superconductors and normal reservoirs starting from
single electron hoppings is presented elsewhere
\cite{sauret_entangler}.
Diagonal elements describe the occupation probabilities of the
configurations
and non-diagonal ones describe the coherences between them.
The latter naturally occur because of the resonant
Andreev processes occurring in the TP sequence, and lead to a
system of Bloch-like equations \cite{cohen}. The derivation is
inspired by the work of Gurvitz \cite{gurvitz,sequential}
who treated the cases of single and double-dot systems.
The generalization starts from an effective Hamiltonian, which is the sum
of
the Andreev pair Hamiltonian and the one-electron injection/detection terms
as in Eq. (\ref{eff}).

Starting with a given initial condition specifying a point of the
sequence,
for instance the state $(10021)$, with a spin $\sigma$ state
in dot $\bf 3$, the Schr\"odinger equation is written for the state at
instant $t$

\begin{equation}
|\Psi (t)\rangle \,=\,\sum_{\mu,n} b_{\mu,n}(t) |\mu,n\rangle
\end{equation}

\noindent
where the index $n$ contains the information
on the quasiparticles which have been transferred (with wave vectors
$k_L,k_R$)
from reservoir $\bf L$ to reservoir
$\bf R$. The reduced density matrix $(\sigma_{\mu \nu})$ involves a trace
of all transition operators $|\mu\rangle\langle\nu|$ on
quasiparticle indices $n$ \cite{gurvitz}.
The index $\mu=b,1,3,A,0,s,t$ runs on the states of the sequence, denoted
for simplicity as $|b\rangle = |10021\rangle$,
$|1\rangle = |10020\rangle$,
$|3\rangle = |02001\rangle$, $|a\rangle = |12001\rangle$, $|0\rangle =
|10001\rangle$, and $|s\rangle$, $|t\rangle$ involved
in the configuration $(10101)$, produced from $(10020)$ as the state
$|\sigma\rangle_1\,|\psi^S\rangle_{23}$.
The decomposition expressed in Eq. (\ref{bell_state_decomposition}) leads
to the singlet state
$|s\rangle = |\psi^S\rangle_{12} |\sigma\rangle_3$ and the triplet
combination
$|t\rangle = (1/\sqrt{3})\sum_{0,+,-}|\psi^{T0,+,-}\rangle_{12}
|\tilde\sigma_{0,+,-}\rangle_3$. The states
$|\tilde\sigma_{0,+,-}\rangle_3$ are obvious notations for the rotated
states appearing in equation (3). Notice that state
$|t\rangle$ does not connect to any other state than $|1\rangle$,
while $|s\rangle$ connects to $|1\rangle$ and $|3\rangle$.

%%%%%%%%%%%%%%%%%%%%%%%%%%%%%%%%%%%%%%%%%%%%%%%%%%%%%%%%%%
\begin{figure} \centerline{\epsfxsize=10cm %\epsfysize=6.5cm
\epsfbox{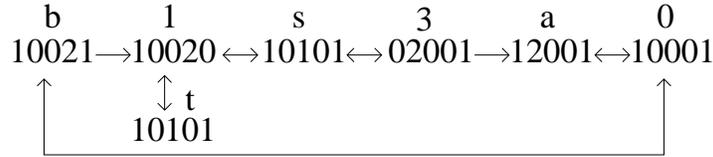}}
%\narrowtext
\vspace{4mm}
\caption{Labeling of the states which enter the Bloch equations. Their
respective
connection is identified by arrows. Single arrows represent irreversible
processes
which involve the transfer of an electron to/from the reservoir.
\label{fig6}}
\end{figure}
%%%%%%%%%%%%%%%%%%%%%%%%%%%%%%%%%%%%%%%%%%%%%%%%%%%%%%%%%%

Let us stress again that spin is perfectly conserved within the
teleportation process. Therefore, a general set of closed Bloch
equations
can be obtained irrespective of the spin
direction along
an arbitrary axis. This of course holds only in absence of any spin 
relaxation or
decoherence. The spin index is thus omitted in the $\sigma_{\mu\nu}$ :

\begin{eqnarray}
\dot\sigma_{\mu \mu} &=&
i\sum_{\nu}\Omega_{\mu \nu}(\sigma_{\mu \nu}-\sigma_{\nu \mu})
- \sum_{\lambda}(\Gamma_{\mu \lambda}\sigma_{\mu \mu}
- \Gamma_{\lambda\mu}\sigma_{\lambda \lambda})\\
\dot\sigma_{\mu \nu} &=& i(E_{\mu}-E_{\nu})\sigma_{\mu \nu} +
i\sum_{\lambda}(\sigma_{\mu\lambda}
\Omega_{\nu\lambda} -\sigma_{\lambda
\nu}\Omega_{\mu \lambda}) -\frac{\sigma_{\mu \nu}}{2}
\sum_{\lambda} (\Gamma_{\mu \lambda} +  \Gamma_{\nu \lambda})
\end{eqnarray}
with $\mu,\nu=a,b,0,1,3,s,t$, and imaginary coefficients ($\Omega_{\mu
\nu}$) on
the right hand side identify coherent processes, while injection/detection
process have real coefficients ($\Gamma_{\mu \nu}$).
For the teleportation cell, according to Fig. \ref{fig6} we have
$\Omega_{a0} =\Omega_{0a} =\Omega_{b0} =\Omega_{0b} =
T_J$, the tunneling rate for Cooper pairs from $\bf a$ to $S$ ($S$
to $\bf b$). $\Omega_{1s} =\Omega_{s1} =-T_A/2$,
$\Omega_{1t} =\Omega_{t1} =-\sqrt{3}T_A/2$,
$\Omega_{3s} =\Omega_{s3} =T_A$ are the Andreev tunneling rates,
$\Gamma_{b1} =\Gamma_R$,
$\Gamma_{3a} =\Gamma_L$, all the other $\Omega_{\mu \nu}$'s and
$\Gamma_{\mu \nu}$'s are zero. The energies $E_{\mu}-E_{\nu}$ of the
Cooper pair transitions
are included for sake of generality, and are set to zero in the resonant
regime.
Note that the Andreev amplitudes $T_A$ are computed the lowest order 
in the Appendix,
yet here these amplitudes are considered non-perturbative, including
all possible round trips between the normal and the superconducting dot.
It is therefore possible to consider the limit of a ``good'' Andreev
contact in what follows.

The full system of Bloch equations involve the populations of the $7$
above states, and the coherences between states $1,3,s,t$
on one hand, and between states $a,b,0$ on the other hand. All the other
coherences are zero since the corresponding states
are coupled by relaxation terms, according to the usual approximation
\cite{cohen}. The Bloch equations are :

\begin{eqnarray}
\dot\sigma_{11}&=&-i\frac{1}{2}T_A(\sigma_{1s}-\sigma_{s1})-i\frac{\sqrt{3}}{2}T_A(\sigma_{1t}-\sigma_{t1})+\Gamma_R\sigma_{bb}
\label{first_bloch}\\
\dot\sigma_{ss}&=&-i\frac{1}{2}T_A(\sigma_{s1}-\sigma_{1s})+iT_A(\sigma_{s3}-\sigma_{3s})
\\
\dot\sigma_{tt}&=&-i\frac{\sqrt{3}}{2}T_A(\sigma_{t1}-\sigma_{1t})
\\
\dot\sigma_{33}&=&iT_A(\sigma_{3s}-\sigma_{s3})-\Gamma_L\sigma_{33}
\\
\dot\sigma_{aa}&=&iT_s(\sigma_{a0}-\sigma_{0a})+\Gamma_L\sigma_{33}
\nonumber\\
\dot\sigma_{00}&=&iT_s(\sigma_{0a}-\sigma_{a0})+iT_s(\sigma_{0b}-\sigma_{b0})
\\
\dot\sigma_{bb}&=&iT_s(\sigma_{b0}-\sigma_{0b})-\Gamma_R\sigma_{bb}
\label{sigma_bb}\\
\dot\sigma_{1s}&=&i(E_1 - E_s) +
i(-\frac{1}{2}T_A\sigma_{11}+T_A\sigma_{13}+\frac{1}{2}T_A\sigma_{ss}+\frac{\sqrt{3}}{2}T_A\sigma_{ts})
\\
\dot\sigma_{1t}&=&i(E_1 - E_t) +
i(-\frac{\sqrt{3}}{2}T_A\sigma_{11}+\frac{1}{2}T_A\sigma_{st}+\frac{\sqrt{3}}{2}T_A\sigma_{tt})
\\
\dot\sigma_{13}&=&i(E_1 - E_3) +
i(T_A\sigma_{1s}+\frac{1}{2}T_A\sigma_{s3}+\frac{\sqrt{3}}{2}T_A\sigma_{t3})-\frac{1}{2}\Gamma_L\sigma_{13}
\\
\dot\sigma_{3s}&=&i(E_3 - E_s) +
i(-\frac{1}{2}T_A\sigma_{31}+T_A\sigma_{33}-T_A\sigma_{ss})-\frac{1}{2}\Gamma_L\sigma_{3s}
\\
\dot\sigma_{3t}&=&i(E_3 - E_t) +
i(-\frac{\sqrt{3}}{2}T_A\sigma_{31}-T_A\sigma_{st})-\frac{1}{2}\Gamma_L\sigma_{3t}
\\
\dot\sigma_{st}&=&i(E_s - E_t) +
i(-\frac{\sqrt{3}}{2}T_A\sigma_{s1}+\frac{1}{2}T_A\sigma_{1t}-T_A\sigma_{3t})
\\
\dot\sigma_{a0}&=&i(E_a - E_0) +
iT_J(\sigma_{ab}+\sigma_{aa}-\sigma_{00})
\\
\dot\sigma_{b0}&=&i(E_b - E_0) +
iT_J(\sigma_{ba}+\sigma_{bb}-\sigma_{00})-\frac{1}{2}\Gamma_R\sigma_{b0}
\\
\dot\sigma_{ab}&=&i(E_a - E_b) +
iT_J(\sigma_{a0}-\sigma_{0b})-\frac{1}{2}\Gamma_R\sigma_{ab}
\label{last_bloch}
\end{eqnarray}

which have to be solved, obviously verifying the normalization 
constraint for the
probabilities

\begin{equation}
\sigma_{11}+\sigma_{ss}+\sigma_{tt}+\sigma_{33}+\sigma_{aa}+\sigma_{00}+\sigma_{bb}=1
\end{equation}

In the above system the equations for $\sigma_{\nu\mu} =
\sigma_{\mu\nu}^*$ are omitted. The stationary solution gives
  the average current flowing from $\bf L$ to $\bf 1$ or
from $\bf 3$ to $\bf R$ for each spin direction.
This current, denoted as the teleportation current, is given by

\begin{equation}
I_{tel,\sigma} = e\Gamma_L \sigma_{33}^{stat}~,
\end{equation}
with $\sigma_{\mu\nu}^{stat}$ the stationary density matrix elements.
Let us first consider the case of resonant Cooper pair transitions. After
a straightforward calculation,
one obtains quite a simple result

\begin{equation}
I_{tel,\sigma}=e\frac{\Gamma_L\Gamma_R}{\alpha\Gamma_L+4\Gamma_R}\frac{T_A^2}{T_A^2+\frac{2\Gamma_L^2\Gamma_R}{\alpha\Gamma_L+4\Gamma_R}}
\label{teleportation_current}
\end{equation}

\noindent
with

\begin{equation}
\alpha = \frac{3}{2} + \frac{1}{2} \frac{\Gamma_R^2}{T_J^2}
\end{equation}

One can show that if by chance the transitions from configuration
$(10020)$ to $(10101)$ and from $(10101)$ to $(02001)$ are not
  exactly resonant,
the above result still holds provided $(10020)$ and $(02001)$ have the
same energy. $T_A$ will then be decreased and can be calculated using the
lowest order perturbation estimate of the Appendix: for instance, an energy
denominator $(\Delta E_{10020}^{10101})^{-1}$ will reflect the suppression
of the Andreev process describing the singlet injection in $\bf 2, 3$.
This situation can be enforced in practice if dots
$\bf 1$ and $\bf 3$ on one hand, and dots $\bf a$ and $\bf b$ on the other
hand, are coupled to the same electrostatic gate in order to preserve 
the symmetry.

Eqs. (\ref{first_bloch}) to  (\ref{last_bloch}) allow to explore the
effect of another detuning effect
such as the energy difference $\delta$ between $(10020)$ and $(02001)$.
One can show that it has a negligible effect provided that $\delta <<
\gamma_{L,R}$.

Equating to zero the term $\dot\sigma_{bb}$ in Eq. (\ref{sigma_bb}) one
easily deduces the Cooper pair current from dot $\bf a$ to dot $\bf b$ :

\begin{equation}
I_{P} = 2eiT_J(\sigma_{b0}^{stat}-\sigma_{0b}^{stat})= 2e\Gamma_R
\sigma_{bb}^{stat} = 2e\Gamma_L \sigma_{33}^{stat}
\end{equation}

One thus finds that $I_P = 2I_{tel,\sigma}$. This relationship expresses a
very fundamental property : each time a spin state
is teleported from $\bf L$ to $\bf R$, a Cooper pair passes from dot $\bf
a$ to dot $\bf b$. This is because the pair transfer
in the S circuit conditions detection. While in the present work
this property can be attributed to the specific model  which we have
chosen for the teleportation cell, and to the parameters
which select the relevant states of the TP cell, the experimental
observation of both currents in such a device, on the
time scale of the teleportation cycles, would
offer a non ambiguous proof of teleportation, as in Ref. \cite{bouw}. 
Instead, here
the equality is demonstrated only on the average.
But it is clear from the operation of the TP sequence, and from the
time-dependent solution of Bloch equations, that it is true at the time
scale of an elementary spin transfer. This
locking of both currents reflects the basic property of TP : splitting a
qubit transfer (electron with its spin) into a
spin transfer (from $\bf 1$ to $\bf 3$) and a classical information
transfer from $\bf a$ to $\bf b$,
which is here nothing but the charge $2e$. This behaves classically owing
to the irreversibility of the transition, driven
by the bias $V$. Notice that the classical signal is sent
automatically and needs no operation external to the circuit.

Here the inter-dot charging energy plays a crucial role because it
conditions the precise sequence. Successive TP cycles follow each other in a
sequential way, overlapping the injection step of cycle $N+1$ and
detection step of cycle $N$ (Fig. \ref{fig4}). In fact, cycle $N+1$ begins
with the injection in $\bf 1$ of an arbitrary spin state. This forces the
Bell measurement of the singlet in $\bf a$ for cycle $N$,
followed by the classical signal and detection of the $N^{th}$ teleported
spin state in $\bf R$. Then $\bf b$ produces the entangled pair for cycle
$N+1$, which is measured in $\bf a$ when the spin state for cycle $N+2$
enters in $\bf 1$, and so on.

Let us insist on the irreversibility of both pair production and
measurement. Although both processes occur between state of identical
energies,
the quantum resonance between $(10020)$, $(10101)$ and $(02001)$ (``odd''
states) "begins" when the previous electron leaves $\bf 3$
and "ends" when the next
one enters $\bf 1$. Then the resonance between  $(12001)$, $(10001)$ and
$(10021)$ (``even'' states) "begins" and "ends" when the electron leaves $\bf
3$.
Change of odd to even states genercally involves a large energy 
change, of a fraction
of $e^2/C$ (see Fig. \ref{fig5}). Thus the phase coherence is lost at 
these transitions,
allowing true transfer of classical information. Yet, the spin coherence
is preserved during each TP cycle.

One sees that the above protocol has the virtue to work in an automatic
way, without any external intervention. It has the great advantage that
its speed is only limited by the Andreev amplitude $T_A$ and
the couplings $\Gamma_{L,R}$, as shown by the average TP current of Eq.
(\ref{teleportation_current}). It seems plausible that $T_A$ can be optimized
by reducing the dimensionality of the superconducting dots, as the
geometrical factor is known to have a reduction effect
in three and two dimensions only.
An estimate of currents in
  the pA range still leads to about $10^7$ TP events per second, a quite
sizable quantity.
One drawback of such a high cycling value is the difficulty of
a time-resolved diagnosis. This will be discussed in a next Section.

\section{Comparison with the quantum optics implementation}

In this work, a primary teleportation diagnosis lies in the nonlocal
transfer of the injected electron spin from $L$ into $R$,
and in the perfect locking of the average TP
current which flows between $L$ and $R$ with the
average pair current in the S-circuit.
This relies on the assumption -- justified by perturbation
theory estimates -- that processes such as successive cotunneling
via the two superconducting dots or such as
Josephson tunneling through the central dot,
which may affect the fidelity of teleportation,
can be neglected: the teleportation channel is then
the dominant one.

However, even if one measures average currents (injection and
detection current, together with the pair current) and if
one finds that these are correlated, this does not
constitute a rigorous proof that we are dealing with
teleportation. In fact, accidental fluctuations in the gate voltages 
for instance may pollute the TP
  process.
To be more precise, the true fingerprint of TP is that
{\it each time} an electron appears
in R with the same spin that was injected from $L$,
a Cooper pair passes almost simultaneously
from $\bf a$ to $\bf b$.
Quantum mechanics described by our microscopic and effective
Hamiltonian confirms explicitly this perfect correlation
of electron and Cooper pair currents at the single
particle/single Cooper pair level.

In similar situations, encountered experimentally first in quantum optics,
such as Bell inequality tests \cite{bell,photons},
a diagnosis which measures correlations between particles
independently of the chosen (classical or quantum) description
of the apparatus is necessary. Such tests of non-locality
allow to rule out a description at the classical level
once the outcome of a specific measurement in known.

In the optics experiment
\cite{bouw}, a coincidence measurement is performed
to isolate the teleportation event.
First, two pairs of photons  are created
at a close time interval with the same laser pulse.
The first pair $\bf 2, 3$ is
the entangled pair which is shared by Alice and Bob
and which is used to build the three-particle state in the Bell state
decomposition.
One of the photons of the second pair $\bf 1, 4$ (the test photon $\bf 1$)
contains the state to be teleported. The other photon of
this same pair ($\bf 4$) serves as a simple trigger to signal when
the propagation of the photon $\bf 1$ has started.
A four-fold coincidence measurement in the trigger detector, in
the two detectors needed to signal the measurement of
a singlet pair $\bf 1$ and $\bf 2$, and in the detector of
the outgoing photon $\bf 3$ allows to confirm the signature of
teleportation.

In nano-circuits, counting single electrons or single Cooper
pairs one by one in a transport experiment still represents a
challenging task. In the present case, they could in principle
be achieved by time resolved capacitive coupling measurements
at the injection and detection location, and on one junction of the
superconducting circuit. Which quantity needs to be measured
to confirm the signature of teleportation ?
Recall that for the theoretical description
of the measurement of entangled states injected from
superconductors, Bell inequality
tests can be envisioned for situations where a stationary current
flows from the superconductors to the detectors, both in
a scattering approach \cite{thierry_Bell} and in sequential
tunneling scenarios. In the former case,
equal time number of particle correlators can be converted into
current--current (noise) cross-correlations.

In the present teleportation scenario, assuming the injected
particle has a definite spin state, a measurement
on both the Cooper pair and on the detected particle is needed.
This involves the knowledge of the quantity
$\langle N_b(t) (\sigma_z)_R(t') \rangle $,
where $N_b(t)$ is the excess Cooper pair arriving in
$\bf b$ and  $(\sigma_z)_R(t')$ is the electron spin
subsequently measured in the detection reservoir $R$
(using a ferromagnetic reservoir).
Note that here $t>t'$ need to belong to the same cycle.
>From the teleportation current result of
Eq. (\ref{teleportation_current}), assuming that the Andreev
tunneling amplitude $T_A$ is comparable to the rates $\Gamma_{L,R}$,
on can estimate the period of each cycle and the condition
on the two times above reads
$(\alpha\Gamma_L+\Gamma_R)/\Gamma_L\Gamma_R>t-t'$.
As in the Bell inequality test
for solid state devices, the above number correlator can readily be
expressed in terms of noise or current-current correlators
at finite frequency:
$\int d\omega e^{i\omega t}\langle  I_P(t) I_{tel}(0) \rangle$,
where $I_{tel}(t)$ is measured using a spin-polarized reservoir.
Note that the choice of having weak injection and escape rates
is fully consistent with the working assumptions for using a
Bloch equation description.

An experimental test of the device would require
to monitor the electron current at the point of injection
and detection, and the Cooper pair current between $a$ and $S$
(or $S$ and $b$),
and to resolve the time correlations \cite{HBT}
between these two currents (in Fig. \ref{fig2}a, such
detectors $D_{L,R,S}$ are sketched).

\section{Conclusion}

To summarize, an electron spin teleportation scheme which employs a
normal/superconducting hybrid nano-device for electrons has been proposed.
Although the overall system contains several elements of controlled size
and nature (superconducting/normal metal/ferromagnetic) which could prove
difficult to integrate, it relies fully on current nano-fabrication
techniques.
The main message of this paper is that teleportation can in principle
be achieved in a hybrid superconductor/normal metal nano-device
operating without time dependent interactions, in the steady state regime,
by simply applying a bias to the device.

We have shown that the teleportation protocol is precisely the same
as in the quantum optics experiment, as only the singlet state of the
Bell decomposition is used. A novelty however for our system is that
the generator for singlet pairs and the detector for such pairs
is a superconductor which either breaks Cooper pairs and distributes
them in the dots or, alternatively  absorbs them, is the same device:
the Andreev-dot entangler \cite{loss}.
We have pointed out that given the microscopic Hamiltonian,
several competing processes are possible in this system.
These can be minimized by adjusting the electrostatic gates
on the dots ands most importantly by working with the condition
of resonant pair transition (singlet pairs of electrons in the
dots and Cooper pairs in the superconducting elements).
This enabled the use of an effective Hamiltonian where the sole
single electron jumps consist of the injection and detection processes
with the reservoirs. Nonetheless the full operation of the device is not
quantum-mechanically coherent, owing to the large energy changes involved in
both the injection/measurement and detection processes.
Finite capacitances of the dots, together with boundary conditions
on the voltage bias and gate voltages, allow to
single out a sequence of successive states where teleportation
takes place. A Bloch equation description provides the derivation of the
teleportation current, which is perfectly locked to the pair
current flowing from one superconducting dot to another.

Limiting factors should also be discussed.
First, although the transport through the dots is described at
the sequential level, it is crucial to maintain spin coherence during the
TP sequence (on a time scale $\sim\hbar/\Gamma_{R,L}$, which turns out to
be ``short'' in practical situations).
This coherence can be destroyed by spin-orbit coupling,
or by collisions with the other electrons within the
dot.  Such spin-flip processes can be minimized by carefully monitoring
the parity of the
occupation number of small enough normal dots.  Second, the present scheme
requires a sufficient
  amplitude $T_A$.  This amplitude is reduced by a
geometrical factor in two and three dimensions when the two N-S tunnel
barriers as spaced farther than a few nanometers
\cite{choi_bruder_loss,loss,falci}.

A more detailed version  of the dynamics of our teleportation  should in
principle also
include unwanted processes. Such processes lead to states which lie
outside the teleportation cycle states, which can be accessed for instance
via
cotunneling transitions. Cotunneling process would then be included by
coherent
couplings of the reduced density matrix elements in the Bloch equation
approach,
and are then likely to reduce the fidelity of teleportation.

How could this device could be implemented experimentally? At the present
time,
the best control of quantum dots is achieved with semiconductor dots
defined by metallic gates. Nevertheless, hybrid,
semiconductor/superconductor
junctions still present technological challenges. Here, reasonably
small barriers have to be achieved between the normal dots and the
superconductor in order to maximize the Andreev injection/absorption rate.
An alternative would be to define
the dots with quasi one--dimensional conductors (nanotubes) placed in
contact with superconducting elements. Indeed, bent, gated, contacted
carbon nanotubes \cite{bezryadin} have demonstrated Coulomb blockade
behavior. Concerning the contact with a superconductor--nanotube
junction, there is some hope that the geometrical constraint
which operates in two and three dimensions is relaxed
\cite{recher,bouchiat}. Also,  the feasibility of
this proposal relies on the control of the single electron
injection and detection process. Efficient spin filters $ L$, $R$,
are already available at low temperatures \cite{spinfilter}.

Finally, it is legitimate to ask about the practical range over which
the electron spin state can be teleported. In the present situation, using
low temperatures, it is likely to be of a few microns. The
proposed setup is generalizable to $2{\cal N} +1$ normal dots,
together with ${\cal N}$ superconducting circuits ($2{\cal N}$
S-dots): TP of a spin state in  dot $\bf 1$ onto dot $2{\cal N}
+1$ can be achieved by a swapping process
\cite{q_information}, thus extending the range of TP.

\acknowledgements

Discussions with V. Bouchiat are gratefully acknowledged.
LEPES is under convention with UJF and INPG, Grenoble. One of us (T.M.)
wishes to thank NTT Basic Research Laboratories for their
hospitality.
%%%%%%%%%%%%%%%%%%%%%%%%%%%%%%%%%%%%%%%%%%%%%%%

\appendix
\section{Perturbative calculation of pair tunneling, cotunneling and
Josephson amplitudes}
\label{tmatrix}

In order to justify our assumptions, we present lowest order perturbative
estimates of the
electron transfer process in the teleportation cell displayed in Figure
\ref{fig3}. For simplicity,
the expressions below are derived with our working assumption that the
charging
energies of the dots are much smaller than the superconducting
gap $\Delta$ (in $\bf a, b, S$). Calculations are performed using the
T-matrix
approximation, in which the ``effective tunneling Hamiltonian'' is
specified by:
\begin{equation}
H_{T}^{eff}=H_T+H_T\sum_n[\frac{1}{i\eta+\varepsilon_i-H_0-H_C}H_T]^n
\end{equation}
with $\varepsilon_i$ the initial energy and the Hamiltonian in the
denominator
excludes all single electron hoppings. $\eta$ is an infinitesimal.
We are interested in events which connect
an initial state $|i\rangle$ (specified by the occupation configurations
$n_1,n_a,n_2,n_b,n_3$ of the dots) which is connected by $H_{T}^{eff}$
to a final state $|f\rangle$ with the same energy.

\subsection{Second order processes}

Assuming no external phase difference across the superconductors, it is
straightforward to derive the
effective amplitudes for pair tunneling ($T_{A}$), cotunneling ($T_C$) and
Josephson tunneling ($T_{J}$) between the superconducting elements, to
second order in the
single electron hopping amplitudes $t_{\mu \nu}$. $T_{J}^{a}$
($T_{J}^{b}$)
involves an intermediate state
with one Bogolubov quasiparticle in $\bf S$ or $\bf a$ ($\bf b$), while
$T_{A}^{a}$ ($T_{A}^{b}$) involves an intermediate state
with one quasiparticle in $\bf a$ ($\bf b$).
The amplitudes $T_{J}^{a,b}$ have
been calculated in Ref. \cite{matveev} and are not reproduced here.
Summing over all possible intermediate states,
the crossed Andreev amplitudes are given by:
\begin{eqnarray}
T_{A}^a\simeq
2\sum_{k} u^{a}_k v^{a}_k t_{1a}t_{2a}/ (i\eta-E_k^{a}) \\
T_{A}^b\simeq
2\sum_{k} u^{b}_k v^{b}_k t_{2b}t_{3b}/ (i\eta-E_k^{b})
\label{Andreev}
\end{eqnarray}

Cotunneling amplitudes are specified in a similar way:
\begin{eqnarray}
T_{C}^a\simeq
\sum_k
t_{1a}t_{a2}(|u^a_k|^2-|v^a_k|^2)/ (i\eta-E_k^{a}) \\
T_{C}^b\simeq
\sum_k
t_{2b}t_{b3}(|u^b_k|^2-|v^b_k|^2)/ (i\eta-E_k^{b})
\label{Cotunneling}
\end{eqnarray}
These amplitudes are reduced from an ideal value of order
$\gamma_{\alpha}$ by the necessary propagation
of the virtual intermediate state quasiparticle between the two junctions,
at a
distance $l$. This
involves a geometrical factor \cite{choi_bruder_loss,loss,falci}, of
order $f(l)=(\lambda_{S}/l)^{2}e^{-l/\pi \xi_{0}}$ in the clean limit
($\lambda_{S}$ is the Fermi length in the superconductor and 
$\xi_{0}$ is the coherence
length).

Note that both the Andreev and the cotunneling amplitudes of Eqs. 
(\ref{Andreev})
and (\ref{Cotunneling}) contain denominators $(i\eta-E_k^{a})$ which 
involve the
superconducting gap. This, however does not imply that the Amplitudes for such
processes are proportional to $\Delta^{-1}$, as one needs to take into account
the energy dependence of $u^{a,b}_k$ and $v^{a,b}_k$ in order to 
compute the sums.
In fact, these processes have either no dependence
on the gap -- the case of the Andreev process -- or a logarithmic dependence
$\ln(W/\Delta)$, where $W$ is the bandwidth of the superconductor in the normal
state -- for the case of cotunneling. Discarding numerical factors of 
order unity,
one finds:
\begin{eqnarray}
T_{A}^a\sim t_{1a}t_{2a}\nu_{S} f(l)\\
T_{A}^b\sim t_{2b}t_{3b}\nu_{S} f(l)\\
T_{C}^a\sim t_{1a}t_{a2}\nu_{S}\ln(W/\Delta) f(l) \\
T_{C}^b\sim t_{2b}t_{b3}\nu_{S} \ln(W/\Delta) f(l)
\label{rough estimates}
\end{eqnarray}
with $\nu_S$ the density of states of the superconductor in the normal state.
For typical physical parameters,
the logarithm is of order one, so the Andreev and the
cotunneling process have a comparable magnitude.

\subsection{fourth order processes}

We consider three types of fourth-order tunneling processes which are
relevant
to our problem: a) teleportation processes; b)
the succession of two cotunneling processes which result in the spin
conserving transfer
of an electron from $\bf 1$ to $\bf 3$ (an unwanted process which pollutes
teleportation);
c) Josephson tunneling from $\bf a$ to $\bf b$ via dot $2$.

For the teleportation process we start with the state ($10020$) as in
Fig. \ref{fig1}. Transferring one electron on $\bf 2$ or $\bf 3$
creates a quasiparticle in $\bf b$ ($10110$ or $10011$), which is
subsequently destroyed when the intermediate state
with the electrons in the normal dots is reached ($10101$).
The final state ($02001$) can be reached via either of the two
states ($11001$ or $01101$) with one quasiparticle in $\bf a$ each.
The off-resonance (OR) teleportation tunneling amplitude then reads:

\begin{equation}
T_{Tel,OR}\simeq -(i\eta-\Delta
E_{10020}^{10101})^{-1}
\sum_k \sum_q
\frac{4 t_{b2} t_{b3} t_{a1}^* t_{a2}^* u^b_k v^b_k
u^{a*}_qv_q^{a*}}{E^b_k E^a_q }
\label{teleportation_amplitude}
\end{equation}

Note that by specifying that the dots level energies in $\bf 1$, $\bf 2$ are
located at
an equal distance but opposite location with respect to the
superconducting chemical
potential, the resonance condition is
enforced: $\Delta E_{10020}^{10101}=0$ (also $\Delta 
E^{02001}_{10101}=0$) and this expression
diverges (except for the presence of $i\eta$). A more careful analysis
would
show that an re-summation of all the terms in the perturbation series
leads to a finite amplitude
in this resonant situation. As specified in the previous section, the
resonant  regime
is the working assumption of our teleportation proposal.

For the successive two cotunneling events, it is possible to write down
a general expression for transitions from the state $n_1n_an_2n_bn_3$
to the final state $(n_1-1)n_an_2n_b(n_3+1)$ via the intermediate state
$(n_1-1)n_a(n_2+1)n_bn_3$:
\begin{equation}
T_{2Cot}\simeq -
(\Delta E_{n_1n_an_2n_bn_3}^{(n_1-1)n_a(n_2+1)n_bn_3})^{-1}
\sum_k \sum_q
\frac{t_{1a} t_{a2} t_{2b} t_{b3} (|u^a_k|^2 - |v^a_k|^2)(u^b_q|^2 -
|v^b_q|^2)}{E^a_k E^b_q }
\label{2cot_amplitude}
\end{equation}
Computing the sums over momentum, this is estimated to be:
\begin{equation}
T_{2Cot}\sim -
(\Delta E_{n_1n_an_2n_bn_3}^{(n_1-1)n_a(n_2+1)n_bn_3})^{-1}
t_{1a} t_{a2} t_{2b} t_{b3} \nu_{S}^2 \ln^2(W/\Delta)
= (\Delta E_{n_1n_an_2n_bn_3}^{(n_1-1)n_a(n_2+1)n_bn_3})^{-1}T_{C}^{a}T_{C}^{b}
\label{2cot_amplitude'}
\end{equation}

\subsection{Estimation of the cotunneling current}

Under the resonance condition, the energy denominator
$E_{n_1n_an_2n_bn_3}^{(n_1-1)n_a(n_2+1)n_bn_3}$
is ``large'' and successive cotunneling can be minimized.
Indeed, using Eq. (\ref{energy_of_config}) and the
definition of the capacitance matrix, one obtains:
\begin{equation}
\Delta E_{n_1n_an_2n_bn_3}^{(n_1-1)n_a(n_2+1)n_bn_3}/e^2
=
{1/2-n_1+Q_1\over C+C_r+C_g}+{1/2+n_2-Q_2\over 2C+C_g}-{n_b-Q_b\over C}
\end{equation}
At this point we use our working assumption that all capacitances
are equal, together with the resonance condition which fixes
$Q_a=Q_b=29/30 $. Furthermore we choose a value of $Q_1$ which is
located well into the stability regions of Fig. \ref{fig5}.
For a given initial configuration, such as
$n_1=1$, $n_2=0=n_3$, $n_a=0$, $n_b=2$, as in the bottom right
corner of Fig. \ref{fig4}, one obtains that:

\begin{equation}
\Delta E_{10020}^{00120}\sim {e^2\over C}
\end{equation}

with a numerical prefactor of order unity.
One can therefore favor processes involving two sequential Andreev
processes
while at the same time reducing the effect of one-electron processes
from $\bf 1$ to $\bf 3$.

Granted, our finding that the intermediate state for the successive
cotunneling process has a ``large'' charging energy $e^2/C$ is
strictly speaking not sufficient to convince oneself that
this process can be ruled out: the typical current associated with
this process needs to be compared to the teleportation current
of Eq. (\ref{teleportation_current}). To estimate this cotunneling
current, we consider $T_{2Cot}$ to be the effective hopping amplitude
between 2 dots, dot $\bf 1$ and dot $\bf 3$, both of which are connected
to reservoirs $\bf L$ and $\bf R$. A Bloch equation approach was
used in Ref. \cite{gurvitz_alone} to describe this situation.
The corresponding current reads:
\begin{equation}
I_{2Cot}=e\frac{\Gamma_L\Gamma_R}{\Gamma_L+\Gamma_R}
\frac{T_{2Cot}^2}{T_{2Cot}^2+\Gamma_L\Gamma_R/4}
\label{gurvitz_current}
\end{equation}
Note that the teleportation current and the cotunneling current
become comparable when $T_A^2, T_{2Cot}^2 >\Gamma_L\Gamma_R$.
The teleportation cell requires the {\it opposite} limit.
Indeed, the injection and evacuation from/to the reservoirs
need to be efficient enough in order to avoid spurious
processes. One is now in a position to justify why cotunneling
can be neglected when compared to the teleportation current.
Assuming $\Gamma_L\sim\Gamma_R$,  $\Gamma_{L,R}\gg T_A,T_{2Cot}$
and neglecting numerical factors, the two currents become:
\begin{eqnarray}
I_{tel,\sigma}&\sim&
e\frac{T_A^2}{\Gamma_{L,R}}
\\
I_{2Cot}&\sim&
e\frac{T_{2Cot}^2}{\Gamma_{L,R}}
\end{eqnarray}
So one simply needs to compare $T_A$ and $T_{2Cot}$.
Using the estimate of Eq. \ref{2cot_amplitude} (assuming all single
electron hoppings to be comparable), the condition
for neglecting the cotunneling current becomes:
\begin{equation}
T_{2Cot}\sim \frac{T_A^2}{\Delta 
E_{n_1n_an_2n_bn_3}^{(n_1-1)n_a(n_2+1)n_bn_3}}\ll  T_A
\end{equation}
that is, the crossed Andreev reflection hopping is required to be much
smaller than the energy difference between the initial
and the intermediate state implied by the sequential cotunneling
process. We have seen in the above that this energy difference is maximized
when the capacitive energy $e^2/C$ is large, this is a confirmation of the
crucial role played by the capacitive couplings in order to select
the proper teleportation sequence.

\subsection{``Other'' (less relevant) fourth order processes}

Note that there are other
higher order cotunneling processes, for instance those which involve 
intermediate
states with two quasiparticles
(one in $\bf a$ and one in $\bf b$ at the same time). One argues here that
they are much weaker because the corresponding energy
denominators all contain the superconducting gap $\Delta^3$.
A typical contribution would read (this time neglecting the charging
energies in front of the gap):
\begin{equation}
T_{2Cot}'\simeq -
\sum_k \sum_q
\frac{t_{1a} t_{a2} t_{2b} t_{b3} (|u^a_k|^2 - |v^a_k|^2)(u^b_q|^2 -
|v^b_q|^2)}{E^a_k E^b_q (E^a_k + E^b_q)}
\label{2cot_amplitude_bis}
\end{equation}
The summations over momenta can be performed, for instance, provided
that, for instance  $(E^a_k + E^b_q)$ is replaced by $E^a_k$ or $E^b_q$.
This yields:
\begin{equation}
\left|T_{2Cot}'\right|
<\left|
\sum_k \sum_q
\frac{t_{1a} t_{a2} t_{2b} t_{b3} (|u^a_k|^2 - |v^a_k|^2)(u^b_q|^2 -
|v^b_q|^2)}{(E^a_k)^2 E^b_q}\right|\sim t_{1a} t_{a2} t_{2b} t_{b3}
\frac{\nu_S^2}{\Delta} \ln{W/\Delta}
\label{2cot_amplitude_majoration}
\end{equation}
One therefore sees that the latter contribution is smaller
than that of Eq. (\ref{2cot_amplitude}) by a factor
$\Delta E_{n_1n_an_2n_bn_3}^{(n_1-1)n_a(n_2+1)n_bn_3}/\Delta$.
Recall that throughout this work, charging energies are assumed
to be smaller than $\Delta$.

Finally, we evaluate some processes contributing to the Josephson
tunneling amplitudes (pair tunneling
from $\bf b$ to $\bf a$ via the central dot $\bf 2$).
The transport properties of a Josephson junction
containing an impurity level (with charging energy)
in the tunnel barrier
has been detailed in Ref. \cite{glazman}.
Two situations
are considered here.
First, assume that dot $\bf 2$ is empty. The transitions
involve an initial state $n_1002n_3$ and a final state $n_1200n_3$.
This transition necessarily involves (at the beginning and at the end)
$2$ intermediate states $n_1011n_3$ and $n_1110n_3$ with one
quasiparticle into $\bf a$ or $\bf b$. The third intermediate state
then has two possibilities: either it involves a quasiparticle in both
$\bf a$ and  $\bf b$ which gives a tunneling amplitude:
\begin{equation}
T_{J2}^{(1)}\approx -\sum_k \sum_q
\frac{t_{b2}^2 t_{2a}^2 u^a_q v^a_q u^{b*}_k v^{b*}_k}
{E^a_q(E^a_q+E^b_k) E^b_k }
\sim -\frac{t_{b2}^2 t_{2a}^2 \nu_S^2}
{\Delta}
\label{josephson1}
\end{equation}
or it implies double occupancy of the central dot:
\begin{equation}
T_{J2}^{(2)}\approx -
(\Delta E_{n_1002n_3}^{n_1020n_3})^{-1}
\sum_k \sum_q
\frac{t_{b2}^2 t_{2a}^2 u^a_q v^a_q u^{b*}_k v^{b*}_k}
{E^a_q  E^b_k }
\sim -\frac{t_{b2}^2 t_{2a}^2 \nu_S^2}
{\Delta E_{n_1002n_3}^{n_1020n_3}}~.
\label{josephson2}
\end{equation}
When one imposes a phase difference between $\bf a$ and $\bf b$,
the fact that the central dot is empty at the initial state
makes this junction a ``0'' junction.
On the other hand, if dot $\bf 2$ is initially occupied, we look for
transitions such as $n_1012n_3\to n_1102n_3\to n_1111n_3
\to n_1201n_3  \to n_1210n_3$:
\begin{equation}
T_{J2}^{(3)}\approx \sum_k \sum_q
\frac{t_{b2}^2 t_{2a}^2 u^a_q v^a_q u^{b*}_k v^{b*}_k}
{E^a_q(E^a_q+E^b_k) E^b_k }\sim -T_{J2}^{(1)}~.
\label{josephson3}
\end{equation}
On the other hand one can have the sequence
$n_1012n_3\to n_1102n_3\to n_12(-1)2n_3
\to n_1201n_3  \to n_1210n_3$ with the amplitude:
\begin{equation}
T_{J2}^{(4)}\approx (\Delta E_{n_1012n_3}^{n_12(-1)2n_3})^{-1}
\sum_k \sum_q
\frac{t_{b2}^2 t_{2a}^2 u^a_q v^a_q u^{b*}_k v^{b*}_k}
{E^a_q E^b_k }\sim \frac{t_{b2}^2 t_{2a}^2 \nu_S^2}
{\Delta E_{n_1012n_3}}~.
\label{josephson4}
\end{equation}
The latter two transitions lead to a $\pi$-junction behavior.
Note that all the Josephson processes either involve three powers of the
superconducting gap in their denominators, or they are proportional to
$\Delta^{-2}$ times the inverse of the Coulomb charging energy for double
occupancy of normal dot $\bf 2$. The choice of normal dots with a small
total capacitance therefore allows to neglect these processes when
compared
to the (resonant) teleportation amplitude.

%%%%%%%%%%%%%%%%%%%%%%%%%%%%%%%%%%%%%%%%%%%%%%%%%

%\end{multicols}

\begin{thebibliography}{99}
%
\bibitem{bennett} C. H. Bennett, G. Brassard, C. Cr\'epeau, R. Josza, 
A. Peres and W. K. Wooters,
  Phys.\ Rev.\ Lett.\ {\bf 70}, 1895 (1993).
%
\bibitem{aspect} A. Aspect, J. Dalibard and G. Roger,
  Phys.\ Rev.\ Lett.\ {\bf 49}, 1804 (1982)
%
\bibitem{mandel_zeilinger} L. Mandel, Rev.  Mod.
Phys. {\bf71}, S274 (1999);
A. Zeilinger, Rev.  Mod. Phys. {\bf71}, S288 (1999).
%
\bibitem{EPR} A. Einstein, B. Podolsky, and N. Rosen, Phys.\ Rev.\ Lett.
{\bf 47}, 777 (1935).
%
\bibitem{zurek} W. K. Wootters and W. H. Zurek, Nature {299}, 802 (1982).
%
\bibitem{bouw} D. Bouwmeester, J.-W. Pan, K. Mattle, M. Eibl, H. 
Weinfurter and A. Zeilinger,
Nature (London) {390}, 575 (1997).
%
\bibitem{autretelep} D. Boschi, S. Branca, F. De Martini, L. Hardy 
and S. Popescu,
Phys. Rev. Lett. {\bf 80}, 1121 (1998);
A. Furusawa, J. Sorensen, S. L. Braunstein,
C. Fuchs, H. J. Kimble, and E. S. Polzik, Science 282, 706 (1998);
S. L. Braunstein and H. J. Kimble, Phys. Rev. Lett. 80, 869 (1998).
%
\bibitem{NMR} M. A. Nielsen, E. Knill and R. Laflamme, Nature 
(London) {\bf 396}, 52 (1998);
I. L. Chuang, L. M. K. Vandersypen, Xinlan-Zhou, D. W. Leung and S. 
Lloyd, Nature {\bf 393}, 143 (1998); J.
A.Jones, M. Mosca
and R. H. Hansen, Nature {\bf 393}, 344 (1998); L. M. K. Vandersypen, 
M. Steffen, G. Breyta, C. S.
Yannoni, R. Cleve and I. L. Chuang, Phys. Rev. Lett. {\bf 85}, 5452 (2000);
L. M. K. Vandersypen, M. Steffen, G. Breyta, C. S.
Yannoni, M. H. Sherwood and I. L. Chuang, Nature {\bf 414}, 883 (2001).
\bibitem{q_information} D. Bouwmeester, A. Ekert and A.
Zeilinger, {\it The Physics of Quantum Information} (Springer-Verlag,
Berlin, 2000).
%
\bibitem{Bennettcorr} C. H. Bennett, D. P. DiVincenzo, J. A. Smolin and W.
K. Wootters,
Phys. Rev. A {\bf 54}, 3824 (1996).
%
\bibitem{larionov_kane} A. A. Larionov, L. E. Fedichkin and K. A. 
Valiev, Nanotechnology {\bf 12},
536 (2001); B. Kane, Phys. Rev. Lett {\bf 90}, 087901 (2003).
%
\bibitem{choi_bruder_loss} M. S. Choi, C. Bruder and D. Loss,
  Phys. Rev. B {\bf 62}, 13569 (2000).
%
\bibitem{loss} P. Recher, E. V. Sukhorukov and D. Loss,
Phys. Rev. B {\bf 63}, 165314 (2001).
%
\bibitem{lesovik_martin_blatter} G. B. Lesovik, T. Martin and
G. Blatter, Eur.\ Phys.\ J.\ B {24}, 287 (2001).
%
\bibitem{DF} G. Deutscher and D. Feinberg, Appl.\ Phys.\ Lett.\
{\bf 76}, 487 (2000).
%
\bibitem{thierry_Bell} N. Chtchelkatchev, G. Blatter, G. Lesovik and T.
Martin,
Phys. Rev. B {\bf 66}, 161320 (2002).
%
\bibitem{excitons} J. A. Reina and N. F. Johnson, Phys. Rev. A {\bf 
63}, 012303 (2000).
%
\bibitem{loss_divincenzo} D. Loss and D. P. DiVincenzo
Phys. Rev. A {\bf 57}, 120 (1998).
%
\bibitem{oliver} W. D. Oliver, F. Yamaguchi and Y. Yamamoto, Phys.\ Rev.\
Lett.\
  {\bf 88}, 037901 (2002).
%
\bibitem{saraga} D. S. Saraga and D. Loss, Phys.
Rev. Lett. {\bf 90}, 166803 (2003).
%
\bibitem{martin_landauer_buttiker} T. Martin and R. Landauer,
Phys.\ Rev.\ B  {\bf 45}, 1742 (1992); M. B\"uttiker, {\it
ibid.} {\bf 46}, 12485 (1992); Phys. Rev. Lett. {\bf 65}, 2901
(1990).
%
\bibitem{HBT} M. Henny, S. Oberholzer, C. Strunk, T. Heinzel, K. 
Ensslin, M. Holland and C. Sch\"onenberger,
  Science {284}, 296 (1999); W. Oliver, J. Kim, R. Liu and Y. Yamamoto,
  {\it ibid}, 299 (1999); T. Martin and R. Landauer,
Phys.\ Rev.\ B  {\bf 45}, 1742 (1992).
%
\bibitem{devoret} M. H. Devoret and H. Grabert, In
{\it Single Charge Tunneling}, H. Grabert and M. H. Devoret eds.
(Plenum, New York 1992); D. Est\`eve, ibid.
%
\bibitem{nous} O. Sauret, D. Feinberg and T. Martin, Eur.
Phys. J. B {\bf 32}, 545 (2003)
%
\bibitem{falci}
G. Falci, D. Feinberg and F. W. J. Hekking,
Europhys.  Lett. {\bf 54}, 255 (2001).
%
\bibitem{epj_feinberg} R. M\'elin and D. Feinberg, Eur. Phys. J. B{\bf 26}, 101
(2002).
%
\bibitem{recher} P. Recher and D. Loss,
Phys. Rev. B {\bf 65}, 165327 (2002).
%
\bibitem{bouchiat} V. Bouchiat, N. Chtchelkatchev, D. Feinberg,
G. Lesovik, T. Martin and J. Torres, Nanotechnology {\bf 14}, 77 (2003).
%
\bibitem{diffusive} For a clean superconductor, $\xi=\xi_0$
is the BCS coherence length. In a dirty superconductor,
with mean-free path $l < \xi_0$, quasiparticle decay on
the coherence length $\xi \sim \sqrt{\xi_0 l}$ and the prefactor 
varies like $R^{-1}$ instead of $R^{-2}$
(D. Feinberg, ArXiv Cond-Mat/0307099; N. M. Chtchelkatchev and M. 
Mar'enko, ArXiv Cond-Mat/0306552).
%
\bibitem{bouw2} D. Bouwmeester, J.-W Pan, H. Weinfurter, and A. Zeilinger,
J. Mod.  Opt.  {\bf 47}, 279 (2000).
%
\bibitem{tinkham} M. Tinkham, {\it Introduction to Superconductivity}
(2nd ed.) (Mc Graw-Hill, New York, 1996).
%
\bibitem{beenakker} C. W. J. Beenakker, Phys. Rev. B {\bf 44}, 1646 (1991).
%
\bibitem{cotunneling} D. V. Averin and Yu.  V. Nazarov, in {\it
Single Charge Tunneling}, H. Grabert and M.H. Devoret eds.
(Plenum, New York 1992).
%
\bibitem{glazman} L. I. Glazman and K. A. Matveev, Pis'ma Zh. Eksp. Teor.
Fiz. {\bf 49}, 570 (1989) [JETP Lett. {\bf 49}, 659 (1989)].
%
\bibitem{matveev} K. A. Matveev, M. Gisself\"alt, L. I. Glazman, M.
Jonson and R. I. Shekhter, Phys. Rev. Lett. {\bf 70}, 2940 (1993).
%
\bibitem{sauret_entangler} O. Sauret, D. Feinberg, T. Martin, in
preparation.
%
\bibitem{cohen} C. Cohen-Tannoudji, J. Dupont-Roc and G. Grynberg,
{\it Atom-Photon Interactions:
Basic Processes and Applications} (Wiley, New York, 1992).
%
\bibitem{gurvitz} S. A. Gurvitz and Ya.  S. Prager, Phys.
Rev. B {\bf 53}, 15932 (1996).
%
\bibitem{sequential} U. Geigenm\"uller
and G. Sch\"on, Europhys. Lett. {10}, 765 (1989);
D. V. Averin, A. N. Korotkov and K. K. Likharev, Phys. Rev. B
{\bf 44}, 6199 (1991).
%
\bibitem{bell} J.S.\ Bell, Rev.\ Mod.\ Phys.\ {\bf 38}, 447 (1966);
J.F.\ Clauser, M. A. Horne, A. Shimony and R. A. Holt,
Phys.\ Rev.\ Lett.\ {\bf 23}, 880 (1969).
%
\bibitem{photons} A. Aspect, J. Dalibard, and G. Roger,
Phys.\ Rev.\ Lett.\ {\bf 49}, 1804 (1982); L. Mandel, Rev.  Mod.
Phys. {\bf71}, S274 (1999);
A. Zeilinger, Rev.  Mod. Phys. {\bf71}, S288 (1999).
%
\bibitem{bezryadin} A. Bezryadin, A.R.M.
Verschueren, S.J. Tans, and C. Dekker,
Phys. Rev. Lett. {\bf 80}, 4036-4039 (1998)
%
\bibitem{spinfilter} R. Fiederling, M. Keim, G. Reuscher, W. Ossau,
G. Schmidt, A. Waag and L. W. Molenkamp, Nature (London) {\bf 402}, 787
(1999); Y. Ohno, D. K. Young, B. Beschoten, F. Matsukura, H. Ohno
and D. D. Awschalom, Nature (London) {\bf 402}, 790 (1999).
%
\bibitem{gurvitz_alone} S. A. Gurvitz, Phys. Rev. B {\bf 57}, 6602 (1998).
%
\end{thebibliography}
\end{document}